\ifdefined\pdfminorversion\pdfminorversion=4\fi
\documentclass[journal]{IEEEtran}
\usepackage[T1]{fontenc}
\usepackage{amsmath,amssymb,amsfonts}
\usepackage{amsthm}
\usepackage{mathrsfs}
\usepackage{graphicx}
\usepackage{tikz}
\usetikzlibrary{patterns}
\usepackage{booktabs}
\usepackage{algpseudocode}
\usepackage{cite}
\usepackage{xcolor}
\usepackage{url}
\urldef{\codeurl}\url{https://github.com/JeanLew01/reachapprox}
\newif\ifshowcomments
\showcommentsfalse

\newif\ifshowedits
\showeditsfalse

\newif\ifrevmark
\revmarkfalse
\DeclareRobustCommand{\rev}[1]{#1}
\newif\iflatestrevmark
\latestrevmarkfalse
\DeclareRobustCommand{\latestrev}[1]{#1}
\allowdisplaybreaks
\newtheoremstyle{ieee}{0pt}{0pt}{\normalfont}{\parindent}{\itshape}{:}{ }{\thmname{#1}\thmnumber{ #2}\thmnote{ (#3)}}
\theoremstyle{ieee}
\newtheorem{assumption}{Assumption}
\newtheorem{definition}{Definition}
\newtheorem{lemma}{Lemma}
\newtheorem{proposition}{Proposition}
\newtheorem{theorem}{Theorem}
\newtheorem{corollary}{Corollary}
\newtheorem{remark}{Remark}

\makeatletter

\renewenvironment{proof}[1][\proofname]{\par\pushQED{\qed}\normalfont\topsep\z@\relax
  \trivlist\item[\hskip\IEEEproofindentspace\itshape #1\@addpunct{:}]\ignorespaces}%
  {\popQED\endtrivlist\@endpefalse}
\makeatother

\newcommand{\mR}{\mathbb{R}}
\newcommand{\cR}{\mathcal{R}}
\newcommand{\cB}{\mathcal{B}}
\newcommand{\cN}{\mathcal{N}}
\newcommand{\cK}{\mathscr{K}}
\newcommand{\Fam}{\mathscr{F}}
\newcommand{\X}{\mathcal{X}}
\newcommand{\supp}{\operatorname{supp}}

\newcommand{\UIC}{\mathrm{UIC}}
\newcommand{\reach}{\operatorname{reach}}
\newcommand{\diam}{\operatorname{diam}}
\newcommand{\Unif}{\mathrm{Unif}}
\newcommand{\inte}{\operatorname{int}}
\newcommand{\conv}{\operatorname{conv}}

\begin{document}
\title{Fundamental Limits of Sampling-Based Reachability: Geometry and Flow Distortion}

\author{Jixian Liu, Ihab Tabbara, Hussein Sibai, and Enrique Mallada
\thanks{This paper extends the CoRL 2026 version~\cite{liu2026limits}. This work was supported by the NSF Global Centers program under Grant No.~2330450, by the DOE Office of Science (ASCR) under Award No.~826565, and by the Amazon AI PhD Fellowship. All code and related materials are available at \codeurl.}%
\thanks{J.~Liu and E.~Mallada are with the Department of Electrical and Computer Engineering, Johns Hopkins University, Baltimore, MD, USA (e-mail: jliu376@jhu.edu; mallada@jhu.edu).}%
\thanks{I.~Tabbara and H.~Sibai are with the Department of Computer Science and Engineering, Washington University in St.\ Louis, St.\ Louis, MO, USA (e-mail: i.k.tabbara@wustl.edu; sibai@wustl.edu).}}

\markboth{Submitted to IEEE Transactions on Automatic Control}
{Liu \MakeLowercase{\textit{et al.}}: Fundamental Limits of Sampling-Based Reachability}

\maketitle

\begin{abstract}
Sampling-based reachability estimates reachable sets from finitely many simulated trajectory endpoints. Existing guarantees often control missed probability mass or approximation \rev{error relative to the convex hull of the reachable set,} leaving the geometric error relative to the reachable set incompletely characterized. We study this error as a problem of support estimation under flow distortion. Our assumptions combine a uniform lower bound on the local volume density of the initial set, a lower bound on the sampling density, and a one-sided Lipschitz constant $\mu$ for the vector field. These conditions admit initial sets with corners and dynamics that may expand or contract. In the fine-accuracy regime, $\tilde{\mathcal O}\bigl(2^ne^{n\mu T}(R/r)^n/(\rho\kappa)\bigr)$ independent endpoint samples suffice for Hausdorff accuracy $r$ using the endpoint cloud. Here, $n$ is the state dimension, $T$ is the horizon, $R$ is the volume-equivalent radius, and $\rho$ and $\kappa$ quantify sampling density and local volume density. For $n\ge2$ and sufficiently small $\kappa$, a minimax lower bound over all estimators based only on independent endpoints matches these dependencies up to a dimension-dependent factor $2\cdot4^n$ and logarithmic factors, within the stated parameter range \latestrev{and when the state domain accommodates the construction}. The exponential burden in dimension, and in the horizon for expanding systems, is therefore intrinsic to this observation model. Planar experiments illustrate the predicted effects of corners and contraction; robot-arm experiments show how empirical accuracy deteriorates with dimension.
\end{abstract}


\section{Introduction}
\label{sec:intro}
\IEEEPARstart{R}{eachability} analysis is a basic tool for certifying the safety of dynamical and learning-enabled systems \cite{holmes2020reachable,meng2022learning,ivanov2021verisig,wu2024verified}. Classical methods based on Hamilton--Jacobi equations \cite{bansal2017hamilton,chen2018hamilton} or on set propagation \cite{girard2005reachability,kurzhanski2000ellipsoidal} give deterministic guarantees but scale poorly with the state dimension, and can be difficult to apply when the dynamics include learned components. Sampling-based methods instead propagate finitely many initial states through the dynamics and build the estimate from the resulting endpoints \cite{liebenwein2018sampling,lew2021sampling,lew2022simple,devonport2021data,selim2022safe,ganai2023iterative,liu2025recurrent}. These methods require only trajectory simulation and are readily parallelized. They form part of a broader effort to exploit sampling and parallel computation in control and optimization, ranging from model predictive path integral control \cite{williams2018information} to diffusion-style trajectory optimizers \cite{pan2024model,xue2025full,xiao2025safediffuser,pan2025sampling}.

\latestrev{Two common types of approximation guarantee} accompany these methods. The first bounds the \latestrev{probability mass or volume} that the estimator fails to cover \cite{liebenwein2018sampling,devonport2021data,devonport2023data}, but may allow a thin or spatially separated part of the reachable set to be missed entirely. The second controls the Hausdorff distance to the convex hull of the reachable set \cite{lew2021sampling,lew2022simple,lew2025convex}, which fills holes and merges components. In the conference version of this work \cite{liu2026limits}, we cast the problem as geometric support estimation under flow distortion. Under positive reach of the closed exterior of the initial set, an $L$-Lipschitz vector field, and a sampling density bounded below, that work establishes finite-sample Hausdorff guarantees and a minimax lower bound of order $(e^{LT}/r)^n$. The lower bound shows that exponential dependence on the dimension $n$ and horizon $T$ can be intrinsic to the endpoint observation model.

Nevertheless, two features of that analysis limit its scope. First, $\reach((\inte S_0)^c)\ge r_0$ requires every point of $S_0$ to lie in a ball of radius $r_0$ inside $S_0$, which is violated as soon as $S_0$ has an outward corner, that is, a nonsmooth boundary point with a tangent cone of aperture smaller than a half-space, such as a polytope vertex or a convex edge; such corners are ubiquitous among the initial sets used in practice. Inward (reflex) corners, by contrast, are admissible, since a locally convex closed exterior does not create the same obstruction. Second, the Lipschitz constant $L$ is a two-sided worst-case quantity: it does not distinguish a contracting from an expanding system, although in the former the reachable set shrinks and estimating it should become easier, not harder, as the horizon grows.

\paragraph{Contributions}
We address both limitations for autonomous systems $\dot x=F(x)$, estimating $S_T=\cR_T(S_0)$ from trajectory endpoints over a family specified by the initial-set scale and geometry, the dynamics, and the sampling law. Our contributions are as follows.
\begin{itemize}
    \item \textbf{Generality.} We identify conditions that convert local probability bounds into geometric coverage: $(\kappa,\lambda)$-standardness of the initial set, which gives a positive lower bound on its local volume density; one-sided Lipschitz continuity of the dynamics with constant $\mu$; and a relative lower bound $\rho$ on the sampling density. These conditions are sufficient for a uniform rate, though not necessary for consistency. The geometric assumption includes every full-dimensional convex polytope, with $\kappa$ given by its smallest vertex solid angle, as well as closures of bounded open sets whose complements have positive reach, while the dynamical assumption accommodates both expansion and contraction.
    \item \textbf{Sample complexity.} We prove a random-covering upper bound: $\tilde{\mathcal O}\bigl((2e^{\mu T}R/r)^n/(\rho\kappa)\bigr)$ endpoint samples suffice for $r$-accurate inner coverage in the fine-accuracy regime, where $R$ is the volume-equivalent radius of the initial set. With an estimator-dependent outer-deviation condition, this yields a Hausdorff guarantee, and contraction provably reduces the sampling burden.
    \item \textbf{Unavoidability.} For $n\ge2$ and \latestrev{sufficiently small $\kappa$}, we establish a minimax lower bound \latestrev{under the scale and domain conditions of Theorem~\ref{thm:lower-std}}: every estimator based only on i.i.d.\ endpoints requires, in the worst case, $\widetilde{\Omega}\bigl((e^{\mu T}R/(2r))^n/(\rho\kappa)\bigr)$ samples. Hence, the dependence on the dimension, time horizon, target accuracy, initial-set scale, sampling density and local volume density is unavoidable for sampling-based reachability from endpoints. In particular, the required number of samples grows exponentially with the dimension $n$, since $e^{\mu T}R/(2r)>8$ wherever the lower bound applies. The upper and lower bounds agree in all these factors up to the constant $2\cdot4^n$ and logarithmic factors.
    \item \textbf{Numerical experiments.} On planar systems with closed-form flows, including a vector field that is one-sided Lipschitz but not globally Lipschitz, the measured Hausdorff error follows the predicted dependence on the corner angle of the initial set, on $\mu$ rather than on the Lipschitz constant, and on the horizon through the factor $e^{\mu T}$. On a multi-link MuJoCo robotic arm, the sample complexity degrades with the state dimension, and adversarial sampling improves accuracy at larger budgets, while the fitted convergence rate still deteriorates with dimension.
\end{itemize}

\paragraph{Notation}
All norms are Euclidean and all logarithms are natural. $|A|$ is Lebesgue measure, $\inte A$ and $\partial A$ are interior and boundary, and $A^c:=\mR^n\setminus A$. $\cB_r(x)$ is the closed ball, $\cB_r:=\cB_r(0)$ and $\omega_n:=|\cB_1|$. $d(x,A):=\inf_{a\in A}\|x-a\|$, $\cB_r(A):=\{x:d(x,A)\le r\}$, $d_H$ is the Hausdorff distance and $\conv$ the convex hull. The covering number $\cN(A,h)$ is the smallest number of closed balls of radius $h$ centered in $A$ whose union contains $A$. \rev{The symbols $\tilde{\mathcal O}$ and $\tilde{\Omega}$ hide logarithmic factors. For $m\ge0$, $\mathbb S^m:=\{z\in\mR^{m+1}:\|z\|=1\}$ is the unit sphere and $\sigma_m$ is its $m$-dimensional surface measure, with $\sigma_0$ the counting measure on $\mathbb S^0=\{-1,1\}$.} For a nonempty closed set $C$, $\reach(C)$ is the supremum of the radii $a$ such that every point at distance less than $a$ from $C$ has a unique nearest point in $C$. For a regular closed set $S=\overline{\inte S}$, its closed exterior is $(\inte S)^c$. Reach assumptions below concern this closed exterior, or the closed complement $A^c$ of an explicitly specified open set $A$.

\section{Preliminaries and Problem Statement}
\label{sec:problem}

We consider an autonomous continuous-time dynamical system $\dot x=F(x)$ \latestrev{on a state domain $\X\subseteq\mR^n$, $n\in \mathbb{N}_{+}$, with a one-sided Lipschitz vector field $F:\X\to\mR^n$}. Given an initial condition $x_0\in\X$, let $\varphi(t,x_0)$ denote the solution at time $t\ge0$, and write $\varphi_T:=\varphi(T,\cdot)$. For a compact initial set $S_0\subseteq\X$ and a time horizon $T\ge0$, the forward reachable set is
\begin{equation*}
    \cR_T(S_0):=\{\varphi(T,x_0)\in\X:\ x_0\in S_0\}.
\end{equation*}
Write $S_T := \cR_{T}(S_0)$. We estimate $S_T$ from endpoints $Y_i:=\varphi_T(X_i)$, $i=1,\dots,N$, where $X_1, \dots, X_N$ are independent and identically distributed (i.i.d.) samples from a probability distribution $P_0$ with $\supp(P_0) = S_0$. Under our standing assumptions, $\varphi_T$ is continuous (Lemma~\ref{lem:flow-bounds}). Consequently, the endpoints are i.i.d.\ samples from the pushforward distribution $P_T:=(\varphi_T)_\#P_0$, with $\supp(P_T)=\varphi_T(S_0)=S_T$ by continuity and compactness. A sampling-based reachable-set estimator is a measurable random closed set $\widehat S_N=\widehat S_N(Y_1,\dots,Y_N)\subseteq\mR^n$. We require both directed distance losses, and hence $d_H(S_T,\widehat S_N)$, to be Borel measurable for every fixed target $S_T$. The Hausdorff loss is $+\infty$ for empty or unbounded estimates. Estimators may depend on the family parameters but, in the endpoint observation model, have no instance-specific information beyond the endpoints.

Probabilistic coverage is a common performance criterion in sampling-based and learning-based reachability analysis. A typical result guarantees $P_T(S_T\setminus\widehat S_N)\le\varepsilon$ with confidence at least $1-\delta$ for an estimate $\widehat S_N$ built from independent endpoints \latestrev{\cite{devonport2023data}}. Such a guarantee bounds the probability mass of the omitted states, but not their distance from the estimated set: a region carrying little mass under $P_T$ may extend far beyond $\widehat S_N$, so a small coverage error can coexist with a large Hausdorff distance. This discrepancy is consequential for safety verification, because rarely reached states may still violate constraints and thereby invalidate a certificate.

To address this gap, we seek finite sample-size bounds that suffice to approximate the reachable set to a specified target Hausdorff distance with high probability, together with minimax lower bounds showing that their dependence on the problem parameters cannot be improved in the worst case, up to dimension-dependent constants and logarithmic factors. Specifically, given an accuracy tolerance $r>0$ and a confidence level $1-\delta$, we characterize conditions on the problem instance and the sample size $N$ that ensure
\begin{equation*}
P_T^N\bigl(d_H(S_T,\widehat S_N)\le r\bigr)\ \ge\ 1-\delta .
\end{equation*}
Unlike a bound on missed probability mass, this guarantee ensures that every point in either set lies within distance $r$ of the other. It controls both omissions that may undermine safety verification and overapproximation that may cause unnecessary conservatism.

A connection between missed probability mass and inner geometric error requires a lower bound on $P_T(\cB_r(x))$ uniform in $x\in S_T$. Indeed, if $d(x,\widehat S_N)>r$, then $\cB_r(x)$ is disjoint from $\widehat S_N$, so its entire mass is omitted. A separate outer-deviation condition is needed to control points of the estimate that lie far from $S_T$.

\begin{figure*}[htbp]
    \centering
    \begin{minipage}[t]{0.325\textwidth}
        \centering
        \includegraphics[width=\linewidth]{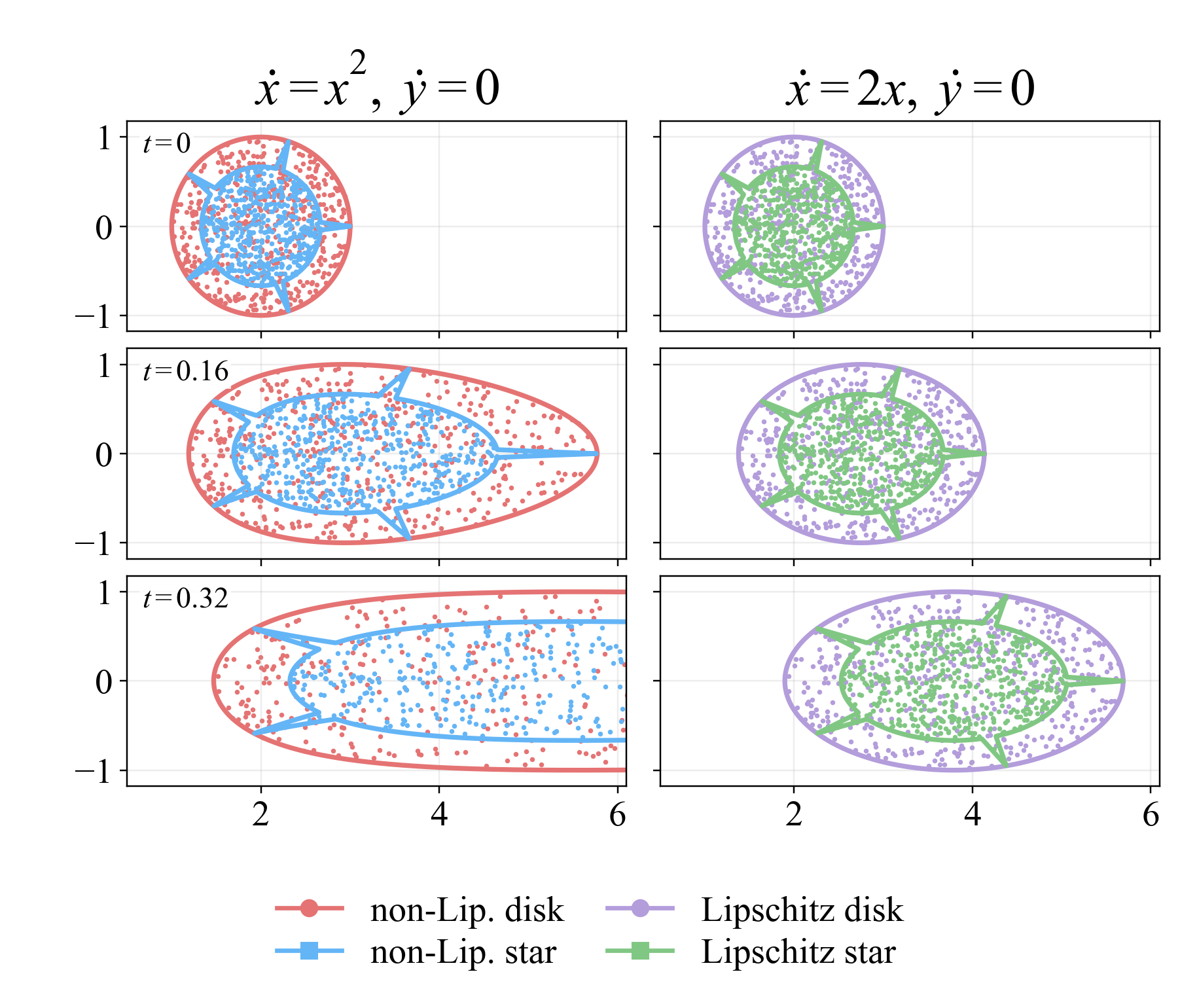}\\
        \footnotesize (a) Reachable sets and samples
    \end{minipage}\hfill
    \begin{minipage}[t]{0.325\textwidth}
        \centering
        \includegraphics[width=\linewidth]{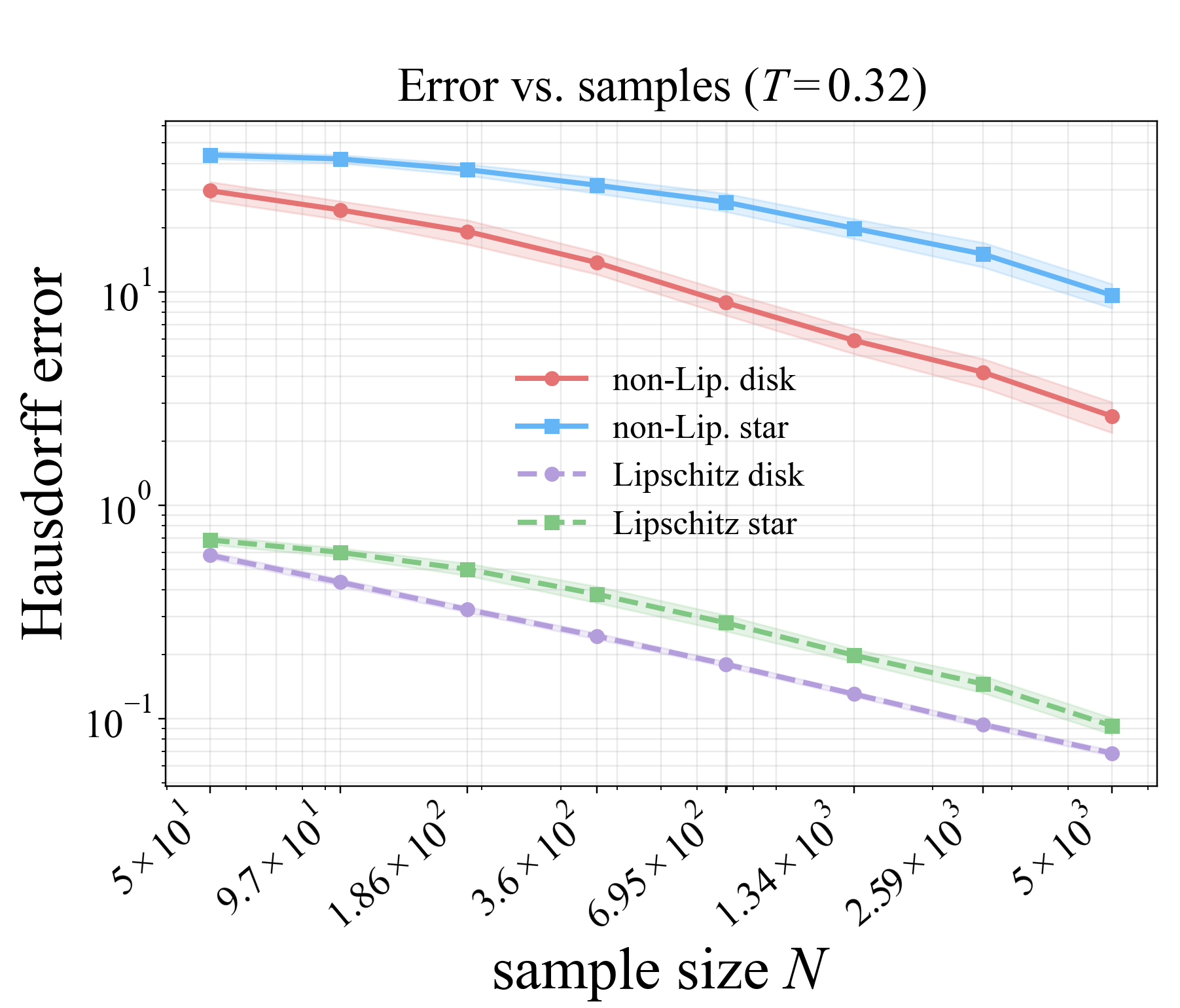}\\
        \footnotesize (b) Error vs.\ sample size
    \end{minipage}\hfill
    \begin{minipage}[t]{0.325\textwidth}
        \centering
        \includegraphics[width=\linewidth]{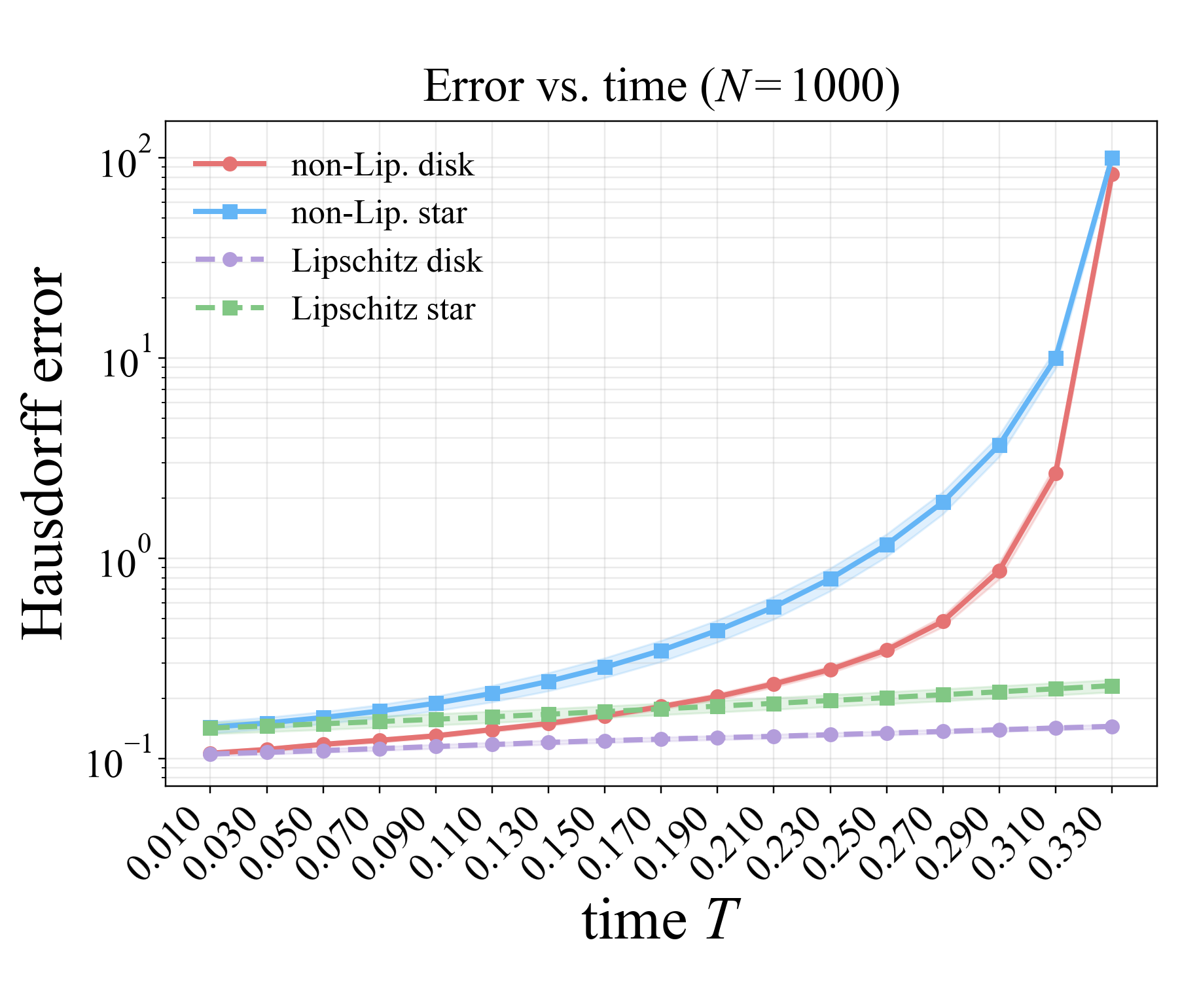}\\
        \footnotesize (c) Error vs.\ horizon
    \end{minipage}
    \caption{Geometry and dynamics. (a) A disk and a nonconvex star of comparable diameter, the latter the union of a circular core of radius $2/3$ of the disk radius and five $15^\circ$ corners, propagated under two fields. Left: $\dot x=x^2$, $\dot y=0$, which is not globally Lipschitz and blows up in finite time; the right tip of the reachable set escapes as $t\uparrow1/3$, and the panel is clipped. Right: the linear field $\dot x=2x$, $\dot y=0$, which is Lipschitz, hence one-sided Lipschitz, and maps the disk to an ellipse. Rows: $t=0$, $0.16$, and $0.32$. Dots: $500$ samples, drawn uniformly by area on each initial set; curves: the exact reachable boundaries. (b) Mean Hausdorff error of the endpoint cloud against $N\in[5\times10^1,5\times10^3]$ at $T=0.32$, and (c) against $T$ at $N=1000$; bands: $95\%$ confidence intervals over $50$ trials. Time changes the distortion of neighborhoods, while the geometry of $S_0$ changes the local area fraction, and hence the local sampling probability, even at a fixed horizon.}
    \label{fig:toy_example}
\end{figure*}

Fig.~\ref{fig:toy_example} illustrates how initial-set
geometry and flow distortion affect the accuracy of sampling-based reachable-set estimation. In panel (a), the same disk and star are propagated under two planar fields. The left column uses the quadratic field $\dot x=x^2$, $\dot y=0$, whose flow exhibits unbounded
stretching as the finite blow-up time is approached.
The right column uses the linear field $\dot x=2x$, $\dot y=0$, under which the disk becomes an ellipse and the reachable set remains bounded at every finite horizon. Comparing the disk and the star highlights
the effect of initial-set geometry, while comparing the
two fields reveals the effect of flow distortion. Panels (b) and (c) quantify the Hausdorff error of the endpoint cloud. At a fixed horizon, the star yields larger
errors than the disk for every tested sample size under
both fields. Under the quadratic field, the star also has a shallower log--log decay slope, so its error decreases more slowly with sample size than that of the disk. With $N$ fixed, the error grows
sharply as $t\uparrow1/3$ under the quadratic field,
whereas under the linear field its logarithm grows approximately linearly with time. These observations highlight how dynamics and geometry
determine local sampling probabilities. A one-sided Lipschitz
bound $\mu$ controls the separation of trajectories by the
factor $e^{\mu T}$, whereas the stretching under the quadratic
field diverges as its blow-up time is approached.
Geometry determines the local volume available for sampling:
the $15^\circ$ corners of the star occupy the fraction
$\frac{15}{360}=\frac{1}{24}$ of sufficiently small disks centered at their tips, whereas the corresponding fraction
at the smooth boundary of the disk tends to
$\frac12$~\cite{liu2026limits}. Under uniform sampling, neighborhoods of these sharp corners therefore carry less probability mass and are harder to resolve. More generally, a sampling-density lower bound prevents regions of positive volume from carrying arbitrarily little mass. These observations motivate conditions on the dynamics, initial-set geometry, and sampling density, which we introduce after stating a well-posedness assumption.

\begin{assumption}[Well-posedness]
\label{ass:well-posed}
For every instance and horizon $T$ considered, $\varphi(t,\cdot)$ is well defined and unique on $S_0$ and $\varphi(t,S_0)\subseteq\X$ for all $t\in[0,T]$.
\end{assumption}

To state the geometric condition, we recall the notion of a standard set \cite{cuevas1997plugin,cuevas2004boundary}.

\begin{definition}[$(\kappa,\lambda)$-standard set]
\label{def:standard}
A compact set $S_0\subset\mR^n$ is \emph{$(\kappa,\lambda)$-standard}, $\kappa\in(0,1]$, $\lambda>0$, if
\begin{equation}
    |\cB_r(x)\cap S_0|\ \ge\ \kappa\,\omega_nr^n\qquad\forall x\in S_0,\ r\in(0,\lambda].
    \label{eq:standard}
\end{equation}
\end{definition}

Standardness is a condition on local volume alone. Proposition~\ref{prop:standard-examples} in Appendix~\ref{app:standard-examples} gives three classes: closures $\overline A$ of open sets with $\reach(A^c)\ge r_0>0$ are $(2^{-n},r_0)$-standard; uniform inner-cone sets with half-angle $\alpha$ are standard with $\kappa=\vartheta_n(\alpha)$, the fraction of the unit ball occupied by the cone; and convex polytopes and their finite unions are standard at an explicit scale, with $\kappa$ determined by their vertex solid angles.

Standardness controls the local thickness of $S_0$ by requiring it to occupy at least a fraction $\kappa$ of every sufficiently small ball centered in $S_0$. It thereby excludes shapes such as outward cusps, whose local volume at the cusp decays faster than $r^n$, from the class for which the uniform rate $r^{-n}$ is proved, although it does not preclude their consistent recovery at a slower rate. One-sided Lipschitz bounds, in turn, control how distances between trajectories evolve.
We then combine these geometric and dynamical conditions with the sampling assumptions to define the following problem family.

\begin{definition}[Problem family]
\label{def:std-family}
Fix constants $R>0$, $\mu\in\mR$, $\kappa\in(0,1]$, $\lambda>0$ and $\rho\in(0,1]$. Let $\Fam^{\rm std}_{R,\mu,\kappa,\lambda,\rho}$ denote the class of triples $(S_0,F,P_0)$ such that:
\begin{enumerate}
    \item $S_0\subset\X$ is a compact set with $|S_0|=\omega_nR^n$ and $|\partial S_0|=0$;
    \item $S_0$ is $(\kappa,\lambda)$-standard;
    \item $F:\X\to\mR^n$ is continuous, satisfies Assumption~\ref{ass:well-posed}, and is one-sided Lipschitz with constant $\mu$, i.e., $\langle F(x)-F(y),x-y\rangle\le\mu\|x-y\|^2$ for all $x,y\in\X$;
    \item $P_0$ is absolutely continuous with respect to the Lebesgue measure on $\mR^n$, with density $p_0$, and satisfies $\supp(P_0)=S_0$ and $p_0(x)\ge\rho/|S_0|$ for a.e.\ $x\in S_0$.
\end{enumerate}
\end{definition}

The one-sided Lipschitz condition in item~3 is weaker than Lipschitz continuity: every $L$-Lipschitz field satisfies it with $\mu=L$, while the cubic field $F(x)=-\|x\|^2x$ satisfies it with $\mu=0$ without being globally Lipschitz (Section~\ref{subsec:exp-osl}).

Three features of Definition~\ref{def:std-family} merit comment. First, for bounded $S_0$ with $|\partial S_0|=0$, taking its closure preserves its volume and absolutely continuous sampling law. If the flow extends continuously to this closure, it also preserves the Hausdorff approximation error. We therefore work with compact $S_0$. The null-boundary condition holds for Lipschitz domains, convex bodies, and finite unions of polytopes. Second, items~1 and~2 impose a quantitative nondegeneracy condition on $S_0$: every ball $\cB_r(x)$ with $x\in S_0$ and $r\le\lambda$ satisfies $ |\cB_r(x)\cap S_0|\ge \kappa\,\omega_n r^n>0.$ Here, $R$ is the radius of a ball with the same volume as $S_0$, while $\rho$ specifies the sampling-density lower bound relative to the uniform distribution on $S_0$. Third, the density bound in item~4 is required only almost everywhere, because $p_0$ is determined by $P_0$ only up to Lebesgue-null sets; in particular, modifying $p_0$ on $\partial S_0$ leaves $P_0$ unchanged. The proofs use the lower bound only in its integrated form: $P_0(A)\ge \frac{\rho}{|S_0|}\,|A\cap S_0|$ for every measurable set $A$. The geometric parameters $(\kappa,\lambda)$ and the dynamical bound $\mu$ determine how this initial density translates into probability mass in neighborhoods of the reachable set. In the next section, we quantify this relationship to establish sample-complexity upper bounds and compare them with minimax lower bounds.

\section{Reachable Set Approximation under Regular Geometry}
\label{sec:regularity}

For the problem family defined above, we study the approximation of the reachable set in Hausdorff distance from finitely many endpoint samples. Our main tool is a local lower bound on endpoint probabilities, derived from the standardness of $S_0$, the lower bound on the initial density, and distortion estimates for the flow induced by the one-sided Lipschitz condition. This estimate yields an inner-coverage sample-complexity bound for any estimator that contains all observed endpoint samples. A full Hausdorff-error guarantee additionally requires an estimator-specific outer-deviation bound. Finally, we show that the $r^{-n}$ scaling is information-theoretically unavoidable, and so is the dependence on $(\rho\kappa)^{-1}$ for every sufficiently small $\kappa$ \latestrev{when the domain accommodates the lower-bound construction}.

\subsection{Local Mass Bounds under Flow Distortion}
\label{subsec:static-mass}

Standardness provides a lower bound on the volume of every small ball centered in $S_0$, which the density assumption in Definition~\ref{def:std-family} converts into a lower bound on its mass under $P_0$. To carry this local-mass bound to the reachable set at time $T$, we do not track the standardness of $S_T$ itself. Instead, we propagate the local thickness of $S_0$ through the flow, whose distortion is controlled by the one-sided Lipschitz constant: the preimage of an $r$-neighborhood centered on the reachable set contains an initial neighborhood of radius $s=e^{-\mu T}r$, whose probability mass is controlled by the local volume bound. This yields, at every scale, the local lower bound on endpoint probabilities and the covering bound used in the sampling argument of the next subsection.

\begin{lemma}[Endpoint local probability lower bound]
\label{lem:endpoint-mass-std}
Let $(S_0,F,P_0)\in\Fam^{\rm std}_{R,\mu,\kappa,\lambda,\rho}$. Then, for every $y\in S_T$ and every $r>0$,
\begin{equation}
    P_T\bigl(\cB_r(y)\cap S_T\bigr)\ \ge\ \rho\,\kappa\Bigl(\frac{\min\{e^{-\mu T}r,\ \lambda\}}{R}\Bigr)^n .
    \label{eq:GammaT-std}
\end{equation}
\end{lemma}

\begin{proof}
   See Appendix~\ref{app:endpoint_mass}.
\end{proof}

The preceding lemma bounds the probability mass of individual neighborhoods from below. To translate this local coverage into a uniform coverage guarantee for $S_T$, we also need to bound the number of such neighborhoods required to cover the reachable set. The following lemma provides this bound.

\begin{lemma}[Covering number of the reachable set at every scale]
\label{lem:covering-std}
Let $(S_0,F,P_0)\in\Fam^{\rm std}_{R,\mu,\kappa,\lambda,\rho}$. Then, for every $r>0$,
\begin{equation}
    \cN(S_T,r)\ \le\ \frac{R^n}{\kappa\,\min\{\frac12e^{-\mu T}r,\ \lambda\}^n}.
    \label{eq:covering-std}
\end{equation}
\end{lemma}

\begin{proof}
   See Appendix~\ref{app:endpoint_mass}.
\end{proof}


\subsection{Sample Complexity Bounds}
\label{sec:bounds}
Lemma~\ref{lem:endpoint-mass-std} shows that every $r$-neighborhood centered on the reachable set carries endpoint probability mass at least $\rho\kappa (e^{-\mu T}r/R)^{n}$ for $r\le e^{\mu T}\lambda$, and at least $\rho\kappa(\lambda/R)^n$ beyond that scale. We use this bound to determine how many endpoints suffice for inner coverage at any accuracy $r>0$. The argument is estimator-agnostic: it requires only that the estimator contain all sampled endpoints. The resulting bound depends on $r$ and $T$ only through the \emph{preimage scale}
\begin{equation*}
    \varrho_T(r):=\min\bigl\{\tfrac12e^{-\mu T}r,\ \lambda\bigr\},
\end{equation*}
the radius of initial neighborhoods $\cB_{\varrho_T(r)}(x)\cap S_0$ that $\varphi_T$ maps into balls of radius $r/2$ (Lemma~\ref{lem:flow-bounds}), capped at the scale $\lambda$ up to which standardness controls the local volume.

\begin{theorem}[Sampling upper bound for inner coverage]
\label{thm:upper-std}
Let $(S_0,F,P_0)\in \Fam^{\rm std}_{R,\mu,\kappa,\lambda,\rho}$,
let $Y_1,\dots,Y_N$ be i.i.d.\ samples from $P_T$, and suppose that the estimator $\widehat S_N$ contains all endpoint samples, i.e., $Y_i\in\widehat S_N$ for $i=1,\dots,N$. Set $D_0:=\diam(S_0)$. For $r>0$ and $\delta\in(0,1)$, if $N$ satisfies any of the following sufficient conditions, then $\sup_{x\in S_T}d(x,\widehat S_N)\le r$ with probability at least $1-\delta$.
\begin{enumerate}
    \item \emph{Fine regime, $0<r\le2e^{\mu T}\lambda$:} $N\ge N_{\mathrm{fine}}(r,\delta)$, where
    \begin{equation}
        {}\hspace*{-1.5em}
        N_{\mathrm{fine}}(r,\delta)
        := \frac{2^{n}e^{n\mu T}R^{n}}{\rho\kappa r^{n}}
        \Bigl[\log\frac{2^{2n}e^{n\mu T}R^{n}}{\kappa r^{n}}
        +\log\frac1\delta\Bigr].
        \label{eq:upper-std-fine}
    \end{equation}
    \item \emph{Saturated regime, $2e^{\mu T}\lambda<r<e^{\mu T}D_0$:} $N\ge N_{\mathrm{sat}}(\delta)$, where
    \begin{equation}
        N_{\mathrm{sat}}(\delta):=
        \frac1{\rho\kappa}\left(\frac{R}{\lambda}\right)^n
        \Bigl[\log\frac{2^nR^n}{\kappa\lambda^n}
        +\log\frac1\delta\Bigr].
        \label{eq:upper-std-sat}
    \end{equation}
    \item \emph{Trivial regime, $r\ge e^{\mu T}D_0$:} one sample suffices, i.e., every $N\ge1$ works.
\end{enumerate}
\end{theorem}

\begin{proof}
    See Appendix~\ref{app:upper_bound}.
\end{proof}

The covering step follows the standard argument of support estimation \cite{cuevas1997plugin,chazal2009sampling}; the new ingredient is the flow-distortion reduction.

\begin{corollary}[Convex initial sets: no saturation]
\label{cor:convex-upper}
Let $(S_0,F,P_0)\in\Fam^{\rm std}_{R,\mu,\kappa,\lambda,\rho}$, assume that $\widehat S_N$ contains all sampled endpoints and that
$S_0$ is convex, and set $D:=\diam(S_0)$. Then $S_0$ is $((R/D)^n,D)$-standard. Hence, Theorem~\ref{thm:upper-std} holds for
every $r>0$ and $\delta\in(0,1)$ provided that
\begin{align}
   & N\ge \frac1\rho\left(\frac{D}{\varrho'_T(r)}\right)^n
    \left[
    n\log\frac{2D}{\varrho'_T(r)}+\log\frac1\delta
    \right],
    \label{eq:upper-convex}
    \\
   & \varrho'_T(r):=\min\{\tfrac12e^{-\mu T}r,D\}.
\end{align}
\end{corollary}

\begin{proof}
See Appendix~\ref{app:upper_bound}.
\end{proof}

In \eqref{eq:upper-convex}, the geometric dependence is expressed through the diameter: substituting $\kappa'=(R/D)^n$ and $\lambda'=D$ into the upper bound gives the prefactor $\rho^{-1}(D/\varrho'_T(r))^n$ and the logarithmic term $n\log(2D/\varrho'_T(r))$.

\begin{remark}[Saturation and convexity]
\label{rem:convex-upper}
The plateau in \eqref{eq:upper-std-sat} reflects the scale $\lambda$ up to which standardness is known. Once $\varrho_T(r)=\lambda$, the local-mass argument no longer improves as $r$ increases. For general standard sets, separated components can produce a genuine plateau, as in Section~\ref{subsec:exp-regimes}. This does not establish a matching minimax lower bound in the saturated regime.

For convex $S_0$, standardness extends to $D=\diam(S_0)$ with constant $(R/D)^n$. Thus, \eqref{eq:upper-convex} improves throughout the nontrivial range $r<e^{\mu T}D$; at $r\ge e^{\mu T}D$, one endpoint already suffices. No saturated interval is needed for this convex-set guarantee.
\end{remark}

\begin{corollary}[Hausdorff bound with outer deviation]
\label{cor:hausdorff-outer-deviation}
Under the assumptions of Theorem~\ref{thm:upper-std}, suppose additionally that the estimator satisfies the outer-deviation condition $\widehat{S}_N \subseteq \cB_\eta(S_T)$ almost surely for some deterministic $\eta\ge0$. Then, with probability at least $1-\delta$, $d_H(S_T,\widehat{S}_N)\le \max\{r,\eta\}$.
\end{corollary}

\begin{proof}
See Appendix~\ref{app:upper_bound}.
\end{proof}

\begin{remark}[Estimator-dependent outer deviation]
\label{rem:estimator-dependent-outer-deviation}
The condition $\widehat S_N\subseteq\cB_\eta(S_T)$ is equivalent to $\sup_{z\in\widehat S_N}d(z,S_T)\le\eta$. In Corollary~\ref{cor:hausdorff-outer-deviation}, Theorem~\ref{thm:upper-std} controls the directed error from $S_T$ to $\widehat S_N$, while $\eta$ controls the opposite direction. For an inner estimator that contains all endpoints, $\eta=0$; for a union of balls of radius $h$ centered at the endpoints, $\eta=h$; and for the convex hull of the endpoints, $\eta=\sup_{z\in\conv(S_T)}d(z,S_T)$.
\end{remark}

We next establish a lower bound that applies to every measurable estimator based on i.i.d.\ endpoint samples. It shows that the dependence on $r^{-n}$, $1/\rho$ and $1/\kappa$ is unavoidable in dimensions $n\ge2$ for every $\kappa\le\kappa_n$ \latestrev{under the domain condition below}, where
\begin{equation}
    \kappa_n:=\frac{|K_n\cap\cB_1|}{\omega_n} \text{ and }
    K_n:=\Bigl\{v\in\mR^n:\ v_1\ge\tfrac{1}{\sqrt2}\,\|v\|\Bigr\}.
    \label{eq:kappa-n}
\end{equation}

The constant $\kappa_n$ is the fraction of the unit ball $\cB_1$ occupied by the circular cone $K_n$ with apex at the center and half-angle $\pi/4$; it is the standardness constant of such a cone at its apex. One has $\kappa_2=\frac14$, $\kappa_3=\frac12\bigl(1-\frac{1}{\sqrt2}\bigr)\approx0.146$ and $\kappa_4\approx0.091$, and $\kappa_n\ge2^{-n}$ for every $n\ge2$, with equality only for $n=2$ (Remark~\ref{rem:lower-kappa}). The range $\kappa\le\kappa_n$ therefore contains every $\kappa\le2^{-n}$, the local volume density that positive reach guarantees.

\begin{theorem}[Minimax lower bound]
\label{thm:lower-std}
Let $n\ge2$, $R>0$, $\mu\in\mR$, $T\ge0$, $\rho\in(0,1]$, $\delta\in(0,\frac12)$, $0<r<e^{\mu T}R/16$, $0<\kappa\le\kappa_n$ and $0<\lambda\le3R/32$. \latestrev{Suppose $\cB_{3R\max\{1,e^{\mu T}\}}(c)\subseteq\X$ for some $c\in\mR^n$}, and define
\begin{equation}
    N_{\mathrm{lb}}(r,\delta)
    :=\frac{1}{2\rho\kappa}
    \Bigl(\frac{e^{\mu T}R}{2r}\Bigr)^n
    \log\frac1{2\delta}.
    \label{eq:lower-std}
\end{equation}
If $N<N_{\mathrm{lb}}(r,\delta)$, then
for every measurable estimator $\widehat S_N$
there exists an instance
$(S_0,F,P_0)\in \Fam^{\rm std}_{R,\mu,\kappa,\lambda,\rho}$
such that
\[
    P_T^N\bigl(
        d_H(\widehat S_N,S_T)>r
    \bigr)>\delta.
\]
\end{theorem}

\begin{proof}
See Appendix~\ref{app:lower_bound}.
\end{proof}

\begin{remark}[Scope of the lower bound]
\label{rem:lower-scope}
Theorem~\ref{thm:lower-std} pertains to the i.i.d.\ endpoint model: its hard instances share the dynamics \latestrev{$F(x)=\mu(x-c)$ on $\X$} and differ only in $S_0$, so knowledge of the dynamics does not distinguish them. Algorithms that know both $S_0$ and $F$, adaptive sampling schemes, and model-aware methods lie outside its scope. The restriction $r<e^{\mu T}R/16$ comes from the two-tip construction, which needs tips separated by more than $2re^{-\mu T}$ inside a body of volume $\omega_nR^n$; whether the bound extends to coarser accuracies, where it would need instances of a different shape, is open.
\end{remark}

The two instances differ only on a sector of radius about $2re^{-\mu T}$ at an apex of local volume fraction $\kappa$, whose probability can be chosen arbitrarily close to $\rho\kappa(2re^{-\mu T}/R)^n$ from above; up to the factor $\frac12\log\frac1{2\delta}$, \eqref{eq:lower-std} is the inverse of that probability. The hard instances lie in the family, and the apex of the kept tip has local volume fraction exactly $\kappa$ (Lemma~\ref{lem:two-tip}(ii)); the bound says nothing about an individual set whose local volume fraction exceeds $\kappa$ everywhere, such as a disk, whose error can be smaller.

Under the restrictions of Theorem~\ref{thm:lower-std} and the fine-regime condition $r\le2e^{\mu T}\lambda$, the upper bound \eqref{eq:upper-std-fine}, applied to the endpoint cloud, and the lower bound \eqref{eq:lower-std} have the same dependence on $e^{\mu T}R/r$, $\rho$ and $\kappa$ for every $\kappa\le\kappa_n$, up to the factor $2\cdot4^n$ and logarithmic factors. This factor is itself exponential in $n$: both bounds exhibit exponential dependence on $n$ and, when $\mu>0$, exponential growth in $T$, but their bases differ by a factor of $4$, so they do not pin down a sharp, dimension-uniform rate. The dependence on $r^{-n}$, $1/\rho$ and $1/\kappa$ is thus unavoidable in the sense of Theorem~\ref{thm:lower-std}, while $e^{n\mu T}$ captures the effects of expansion and contraction. Table~\ref{tab:upper-lower} summarizes the comparison.

\begin{table}[htbp]
\centering
\caption{\latestrev{Sample-complexity bounds in the common regime specified above.}}
\label{tab:upper-lower}
\small
\renewcommand{\arraystretch}{1.3}
\begin{tabular}{@{}lll@{}}
\toprule
Bound & Range of $\kappa$ & Sample complexity\\
\midrule
Upper bound
& $0<\kappa\le1$
& $\tilde{\mathcal O}\Bigl(
    \dfrac{1}{\rho\kappa}
    \Bigl(\dfrac{2e^{\mu T}R}{r}\Bigr)^n
  \Bigr)$
\\[8pt]
Lower bound
& $0<\kappa\le\kappa_n$
& $\Omega\Bigl(
    \dfrac{1}{\rho\kappa}
    \Bigl(\dfrac{e^{\mu T}R}{2r}\Bigr)^n
    \log\dfrac1{2\delta}
  \Bigr)$
\\
\bottomrule
\end{tabular}
\end{table}

\section{Numerical Experiments}
\label{sec:exp}
Our experiments isolate the two ingredients of Section~\ref{sec:regularity} that go beyond Liu et al.~\cite{liu2026limits}: the factor $1/\kappa$, which stems from a regularity condition on the local volume of $S_0$ rather than on the shape of its boundary, a condition weaker than positive reach of the closed exterior; and the one-sided Lipschitz constant $\mu$, which, unlike the Lipschitz constant $L\ge0$, can be negative and therefore captures contraction. Sections~\ref{subsec:exp-corners} and~\ref{subsec:exp-osl} test each ingredient separately.

We then test the sample-complexity bounds that combine them. The upper bound of Theorem~\ref{thm:upper-std} and the lower bound of Theorem~\ref{thm:lower-std} depend on $\kappa$ and $\mu$ through $e^{n\mu T} (R/r)^n/(\rho\kappa)$. Section~\ref{subsec:exp-regimes} compares both bounds with the empirical sample complexity along the horizon.

These three experiments use planar systems with closed-form flows and explicit geometric parameters. A fourth experiment, on a multi-link robot arm, examines the dependence on the state dimension under uniform and adversarial sampling.

\paragraph{Setup of the planar experiments}
The initial sets have area $\pi$ and are sampled uniformly, so $R=1$ and $\rho=1$, and we take $\delta=0.05$. We take the endpoint cloud $\widehat S_N=\{Y_1,\dots,Y_N\}$ as the estimator. It has $\eta=0$ in Corollary~\ref{cor:hausdorff-outer-deviation}, so that $d_H(S_T,\widehat S_N)=\sup_{x\in S_T}d(x,\widehat S_N)$, and every other estimator that contains the samples has an inner error at most as large. We approximate this distance using the endpoint Voronoi diagram and a discretization of $\partial S_T$. A maximum of the nearest-sample distance occurs either at an interior Voronoi vertex or on the boundary. For these smooth, invertible planar flows, membership is tested using $\varphi_T^{-1}$ and the boundary is propagated from $\partial S_0$. The initial boundary spacing is at most $5\times10^{-4}$. For exact boundary samples, the resulting discretization error is at most $e^{\mu T}\times5\times10^{-4}$; the interior Voronoi contribution is exact in exact arithmetic. Floating-point and polygonal boundary-approximation errors are additional numerical errors. Every condition is repeated over $200$ independent trials. Three of the four vector fields are the linear fields
\begin{equation}
    F_\mu(x)=\begin{bmatrix}\mu&-\omega\\ \omega&\mu\end{bmatrix}x,\qquad\omega=4,
    \label{eq:spiral}
\end{equation}
whose flow $\varphi_T(x)=e^{\mu T}\,\mathrm{Rot}(\omega T)\,x$ is a rotation composed with a scaling. \rev{Here $\mathrm{Rot}(\theta)$ denotes the counterclockwise planar rotation through angle $\theta$.} The one-sided Lipschitz constant of $F_\mu$ is $\mu$, of either sign, whereas its Lipschitz constant is $L=(\mu^2+\omega^2)^{1/2}\ge4$.

\subsection{Dependence on the Corner Angle}
\label{subsec:exp-corners}
\begin{figure*}[htbp]
    \centering
    \setlength{\abovecaptionskip}{2pt}
    \setlength{\belowcaptionskip}{0pt}
    \includegraphics[width=\textwidth]{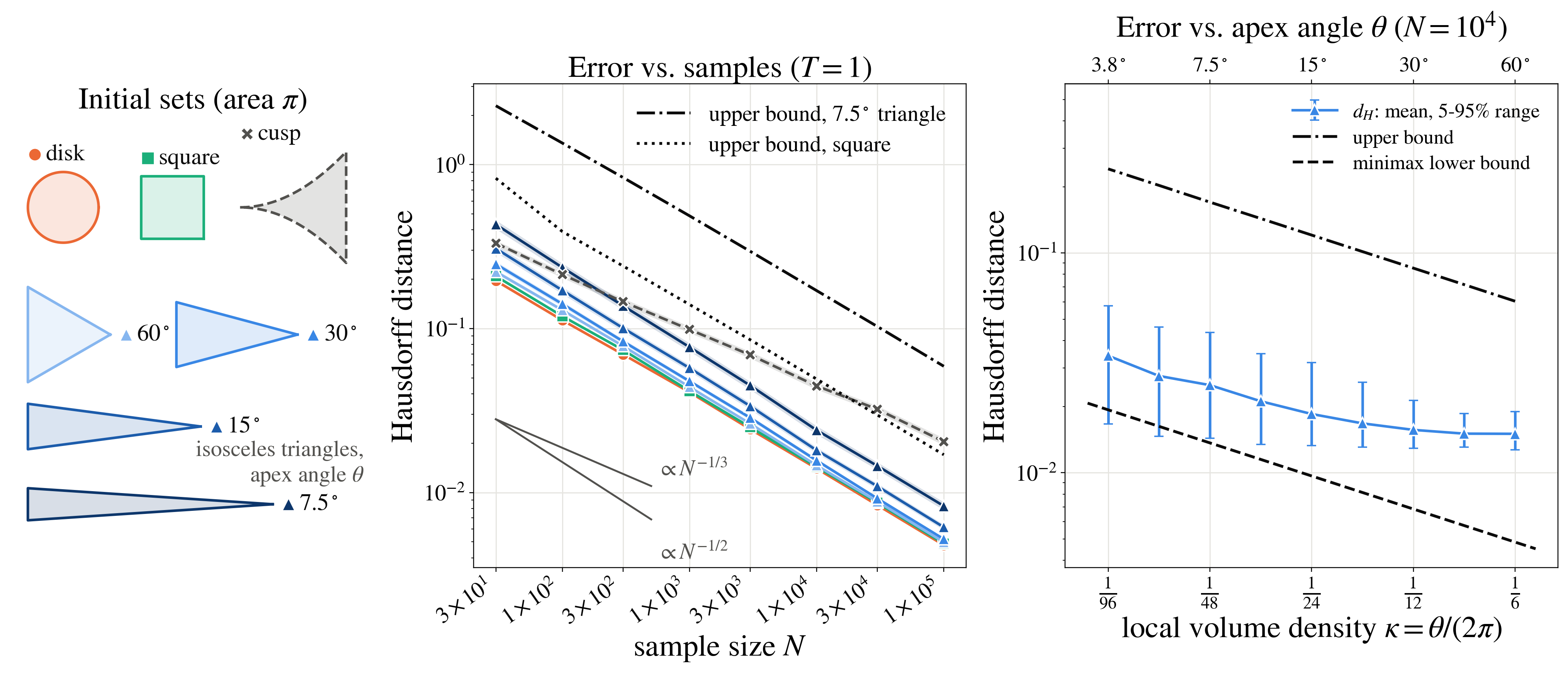}
    \caption{Effect of the corner angle under the spiral sink $F_{-1}$ at $T=1$. Left: equal-area initial sets. Middle: mean Hausdorff error versus $N$, with $95\%$ confidence bands and the upper bounds solved for $r$. Right: mean error and empirical $5$--$95\%$ ranges at $N=10^4$, together with the upper bound and the minimax lower bound \eqref{eq:lower-std} solved for $r$ at $\delta=0.05$. The lower curve is a worst-case bound over the admissible family of instances, not a bound on the mean or quantile of the error for each plotted triangle.}
    \label{fig:exp-corners}
\end{figure*}

We take $\mu=-1$ and $T=1$ in \eqref{eq:spiral} and vary the initial set: the unit disk, for which $\reach((\inte S_0)^c)=1$; a square; isosceles triangles with apex angle $\theta\in\{60^\circ,30^\circ,15^\circ,7.5^\circ\}$, for which $\kappa=\theta/(2\pi)$ by Proposition~\ref{prop:standard-examples}; and a translated copy of the closed cusp $\{0\le u\le3,\ |v|\le\frac\pi2(u/3)^2\}$, which is not standard because its area in a disk of radius $r$ centered at the origin scales as $r^3$. The middle panel of Fig.~\ref{fig:exp-corners} shows the error for $N$ from $30$ to $10^5$. The curves of all standard sets are parallel: over $N\ge10^3$ their fitted log--log slopes lie between $-0.46$ and $-0.49$, which is the rate $N^{-1/n}$ of \eqref{eq:upper-std-fine} up to its logarithm, and the corners of the square and of the triangles, which force the reach of their closed exteriors to vanish, do not degrade it. The cusp decays with slope $-0.34$, consistent with the rate $N^{-1/3}$ suggested by its local area scaling. These observations are consistent with the role of standardness in preserving the exponent despite corners. A corner does change the constant: the curves are ordered by $\kappa$, and at $N=10^5$ the error of the $7.5^\circ$ triangle is $1.7$ times that of the disk.

The right panel isolates the corner effect at $N=10^4$ by decreasing $\theta$ from $60^\circ$ to $3.75^\circ$, or $\kappa$ from $1/6$ to $1/96$. It shows the mean error and empirical $5$--$95\%$ ranges, together with the upper bound and the minimax lower bound \eqref{eq:lower-std}, both solved for $r$ at $\delta=0.05$. The lower curve is
\begin{equation*}
    r_{\mathrm{lb}}(N,\delta)
    =\frac{e^{\mu T}R}{2}
    \left(\frac{\log(1/(2\delta))}{2\rho\kappa N}\right)^{1/n}.
\end{equation*}
The plotted range satisfies $r_{\mathrm{lb}}<e^{\mu T}R/16$ and $\kappa\le\kappa_2=1/4$. The lower bound concerns the family with standardness scale $\lambda_{\rm lb}=\min\{\lambda,3R/32\}$, rather than either the mean or the quantile of the error for a fixed triangle.

For $\kappa\le1/48$, the mean Hausdorff error is within approximately $7\%$ of the mean distance $d_\ast$ from the propagated apex to the endpoint cloud. At radii $s$ for which the pulled-back neighborhood remains within the apex cone, uniform sampling gives $\mathbb P(d_\ast>s)=[1-\kappa(e^{-\mu T}s/R)^2]^N$. Including the four thinner triangles down to $\kappa=1/384$ omitted from the figure, the fitted exponents of the mean and empirical $95\%$ quantile with respect to $\kappa$ are $-0.49$ and $-0.51$. For wider angles, the interior and edges contribute more strongly and the mean error levels off. Over the plotted range, the empirical $95\%$ quantile is a factor of $3.2$ to $4.4$ below the upper bound and $2.8$ to $3.9$ above the minimax lower curve. For thin triangles, both the measured error and the bounds exhibit the $\kappa^{-1/2}$ dependence.

\subsection{One-Sided versus Two-Sided Lipschitz Constants}
\label{subsec:exp-osl}

\begin{figure*}[t]
    \centering
    \includegraphics[width=\textwidth]{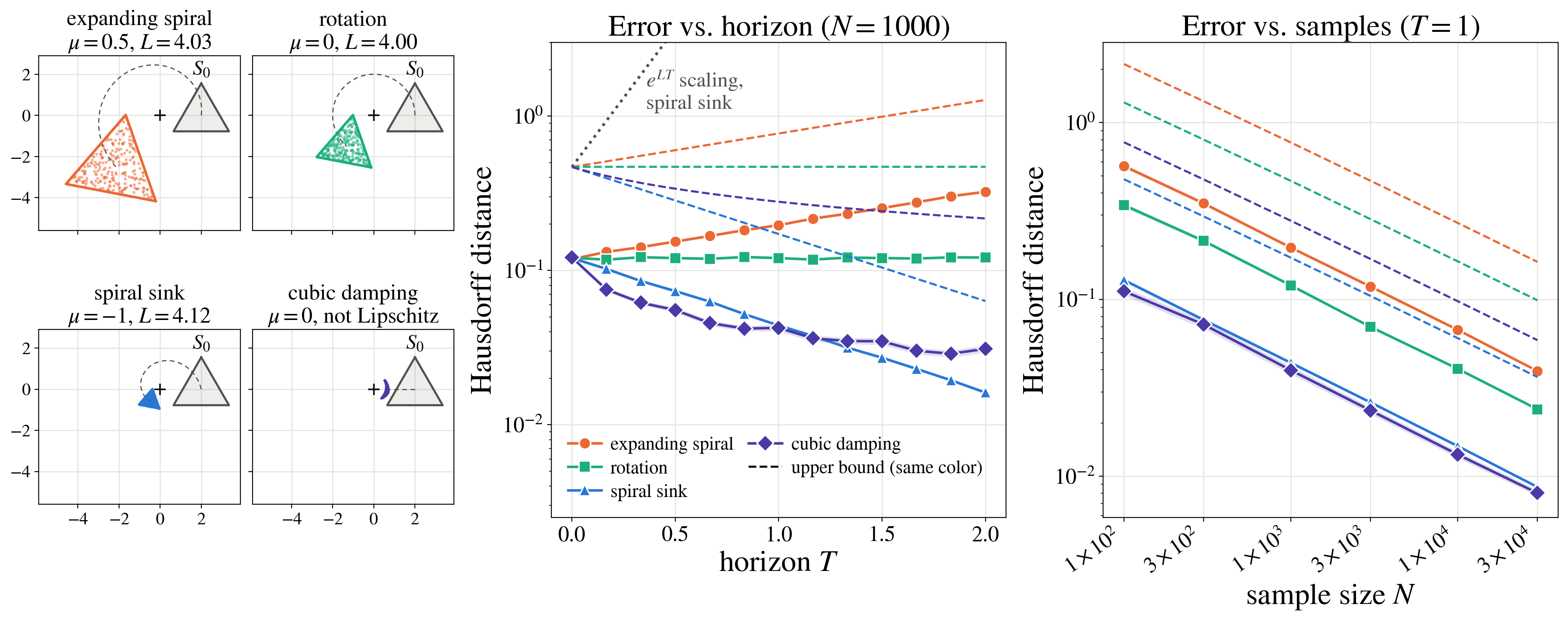}
    \caption{\latestrev{Four vector fields on the equilateral triangle of area $\pi$. Left: $S_0$, $S_T$ at $T=1$ with $250$ endpoints, and the centroid trajectory. Middle: mean Hausdorff error against $T$ at $N=1000$. Right: error against $N$ at $T=1$. Bands: 95\% confidence intervals. Dashed: the upper bound of Theorem~\ref{thm:upper-std} and Corollary~\ref{cor:convex-upper} solved for $r$, with $e^{\mu T}$ for the linear fields and a Lipschitz bound for $\varphi_T$ on $S_0$ for the cubic field. Dotted: the growth $e^{LT}$ that a two-sided Lipschitz bound predicts for the spiral sink.}}
    \label{fig:exp-osl}
\end{figure*}

\begin{table}[htbp]
\centering
\caption{Horizon dependence of the Hausdorff error under four planar vector fields.}
\label{tab:exp-osl}
\small
\renewcommand{\arraystretch}{1.15}
\begin{tabular}{@{}lrrrrrr@{}}
\toprule
Field & $\mu$ & $L$ & rate & ratio & $e^{2\mu}$ & $e^{2L}$\\
\midrule
$F_{0.5}$ & $0.5$ & $4.03$ & $0.50$ & $2.74$ & $2.72$ & $3.2\times10^{3}$\\
$F_{0}$ & $0$ & $4.00$ & $0.00$ & $1.00$ & $1$ & $3.0\times10^{3}$\\
$F_{-1}$ & $-1$ & $4.12$ & $-1.00$ & $0.134$ & $0.135$ & $3.8\times10^{3}$\\
cubic & $0$ & $35.4$ & --- & $0.256$ & $1$ & $5.8\times10^{30}$\\
\bottomrule
\end{tabular}
\end{table}

In Table~\ref{tab:exp-osl}, the rate is the least-squares slope of the logarithm of the mean error over $T\in[0,2]$, and the ratio is the mean error at $T=2$ divided by that at $T=0$.

The initial set is the equilateral triangle of area $\pi$ centered at $(2,0)$, for which $(\kappa,\lambda)=(1/6,0.778)$ and whose closed exterior has reach zero. We compare $F_\mu$ in \eqref{eq:spiral} for $\mu\in\{0.5,0,-1\}$ with the cubic damping field $F(x)=-\|x\|^2x$, whose flow is $\varphi_T(x)=x/(1+2T\|x\|^2)^{1/2}$. This field is the negative gradient of a convex function, so $\mu=0$, but it is not globally Lipschitz; on the ball that contains the trajectories its Lipschitz constant is $L=35.4$.

For the three linear fields the error is exponential in $T$ with fitted rates $0.50$, $0.00$ and $-1.00$, that is, with rate $\mu$, although $L\approx4$ for all of them (Fig.~\ref{fig:exp-osl}, middle, and Table~\ref{tab:exp-osl}). A bound stated in terms of $L$ predicts the growth $e^{LT}$ for all three, a factor between $3.0\times10^3$ and $3.8\times10^3$ at $T=2$, even for the sink, whose error actually decreases by the factor $0.134\approx e^{-2}$. The upper bound of Theorem~\ref{thm:upper-std} follows each curve and stays a factor of about $3$ above its $95\%$ quantile over the whole horizon. For the cubic field, $\mu=0$ gives a horizon-independent upper bound, even though no global Lipschitz constant exists. On the convex set $S_0$, the Jacobian of $\varphi_T$ has tangential and radial eigenvalues $(1+2T\|x\|^2)^{-1/2}$ and $(1+2T\|x\|^2)^{-3/2}$. Thus $L_T:=(1+2Td_0^2)^{-1/2}$, with $d_0:=d(0,S_0)\approx0.954$, is a Lipschitz bound. Substituting $L_T$ for $e^{\mu T}$ is justified by Remark~\ref{rem:lipschitz-map} and gives the decreasing dashed curve; the measured error decreases faster still, by the factor $0.256$ at $T=2$ against $0.464$, because the flow contracts the radial direction more strongly than the tangential one. The sweep over $N$ at $T=1$ (right) shows that the dynamics shift the curves and leave their fitted slopes similar (approximately $-0.47$ for all four fields); for the three linear fields the shift is the factor $e^{\mu T}$.

\subsection{Sample Complexity under Contraction}
\label{subsec:exp-regimes}

\begin{figure*}[t]
    \centering
    \includegraphics[width=\textwidth]{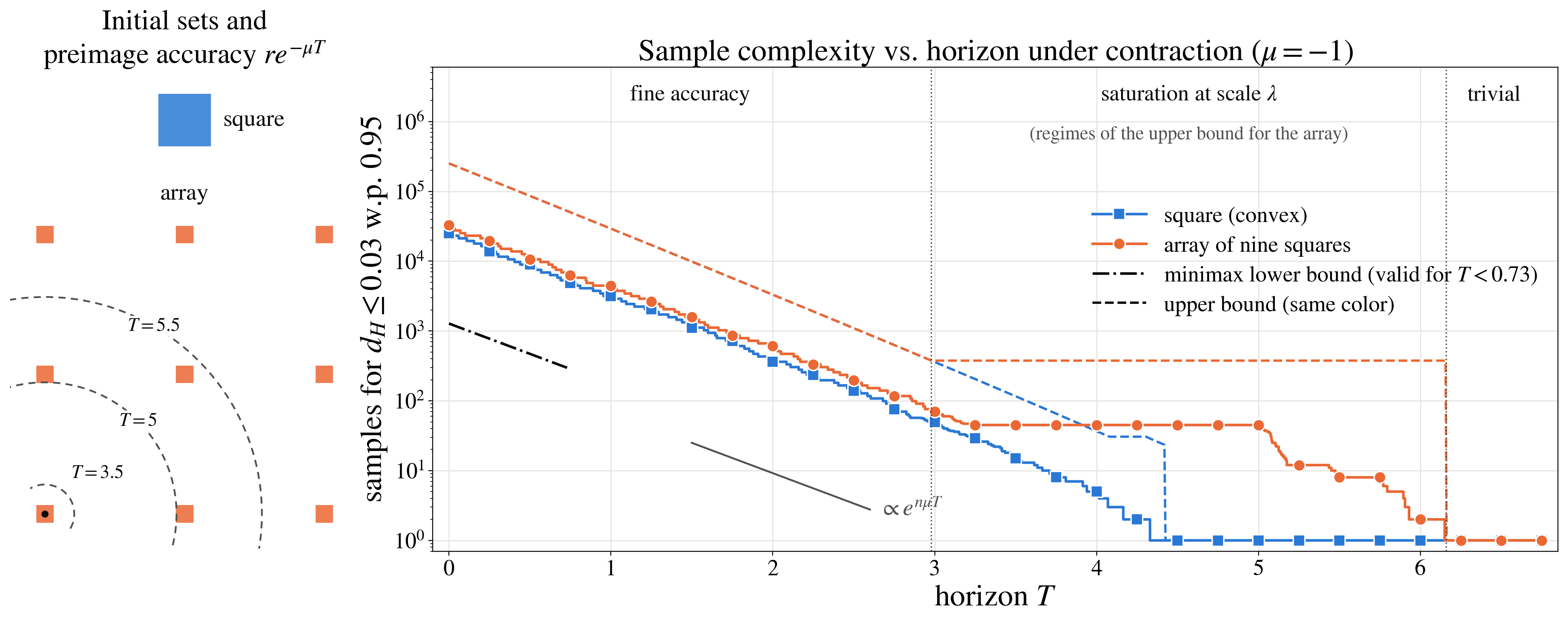}
    \caption{Sample complexity against the horizon under the spiral sink $F_{-1}$. Left: the two initial sets to scale; the dashed arcs have radius $re^{-\mu T}$, $r=0.03$, around a point of a corner piece. Right: samples needed for $d_H(S_T,\widehat S_N)\le r$ in $95\%$ of $200$ trials, on a grid of step $0.005$ in $T$ (solid steps) and at the horizons $T\in\{0,0.25,\dots\}$ (markers); the upper bound (dashed); and the minimax lower bound \eqref{eq:lower-std} (dash-dotted), drawn only where it applies, $r<e^{\mu T}R/16$, that is, $T<0.73$. Vertical lines mark the horizons where the preimage scale reaches $\lambda$ and where the one-sample bound begins.}
    \label{fig:exp-regimes}
\end{figure*}

We keep $F_{-1}$ and fix the accuracy $r=0.03$. Along one i.i.d.\ sequence per trial we record the first $N$ for which $d_H(S_T,\{Y_1,\dots,Y_N\})\le r$, with all integer budgets up to $64$ and a geometric grid of ratio $2^{1/8}$ thereafter, and use its empirical $95\%$ quantile to estimate sample complexity. Theorem~\ref{thm:upper-std} bounds the population quantile; the empirical quantile also reflects finite-trial variability and the grid in $N$. Since $\varphi_T$ is a rotation composed with the scaling $e^{\mu T}$, $d_H(S_T,\{Y_i\}_{i\le N})=e^{\mu T}d_H(S_0,\{X_i\}_{i\le N})$, so one error curve per trial yields this number for every horizon, and we evaluate it on a grid of step $0.005$ in $T$. We compare two initial sets (Fig.~\ref{fig:exp-regimes}, left): a square, which is convex with $(\kappa,\lambda)=(1/4,0.886)$, and an array of nine squares of side $0.59$ on a lattice of pitch $4.73$, which is $(1/4,0.295)$-standard by Corollary~\ref{cor:simplex}(ii) and is an instance of a disconnected initial set.

For $T\le2.75$, $r\le2e^{\mu T}\lambda$ for both sets, and their sample complexity decays like $e^{n\mu T}$: the fitted rates are $-2.08$ for the square and $-2.02$ for the array against $n\mu=-2$, from about $3\times10^4$ samples at $T=0$ to about $10^2$ at $T=2.75$. The upper bound is parallel to the data and a factor $5$ to $11$ above them; the minimax lower bound \eqref{eq:lower-std} with $\kappa=\frac14$, which applies only while $r<e^{\mu T}R/16$, that is, for $T<0.73$, lies a factor of $18$ to $26$ below the empirical curves on that interval. The lower curve concerns the larger family with standardness scale $\lambda_{\rm lb}=\min\{\lambda,3R/32\}$; it is not a lower bound for either fixed initial set or for the smaller family with its original, larger $\lambda$. Contraction reduces the empirical sampling burden at a rate consistent with Theorem~\ref{thm:upper-std}.

At $T=2.98$, the preimage scale $\frac12e^{-\mu T}r$ of the array reaches $\lambda$, and its bound saturates at $376$ samples. The empirical sample complexity saturates as well and equals $45$ for every $T\in[3.25,5]$. To explain this plateau, let $\ell$ be the side of a piece and $p=8\ell$ be the pitch. For $T\in[3.5,4.995]$, the accuracy pulled back to the initial set, $re^{-\mu T}\in[0.99,4.43]$, exceeds the diagonal $0.84$ of a piece, so a sample inside a piece covers the whole piece; it is also below $p-\ell/2\approx4.431$, the distance from the center of any piece to the nearest other piece. On this interval, therefore, the reachable set is covered exactly when every piece has been hit, and the $95\%$ quantile of the corresponding coupon-collector time for nine equally likely pieces is $44$. At $T=5$ the pulled-back accuracy is $4.452>4.431$, so this sufficient condition for the coupon-collector equivalence no longer holds beyond $T\approx4.995$. The square, in contrast, keeps improving and needs a single sample from $T=4.33$ on, consistent with Corollary~\ref{cor:convex-upper} and case (3) of Theorem~\ref{thm:upper-std}, which guarantee one sample for $T\ge\log(\diam(S_0)/r)=4.43$. For the array the one-sample bound starts at $T=6.16$ in the bound and at $T=6.15$ in the data. Between $T=5$ and $T=6.16$ a sample starts to cover neighboring pieces and the empirical sample complexity falls below the plateau in steps, which \eqref{eq:upper-std-sat} does not follow; this gap between the empirical plateau and the saturated bound is listed among the open problems of Section~\ref{sec:conclusion}.

\subsection{Dependence on the State Dimension}
\label{subsec:exp-arm}

\begin{figure*}[t]
    \centering
    \includegraphics[width=\textwidth]{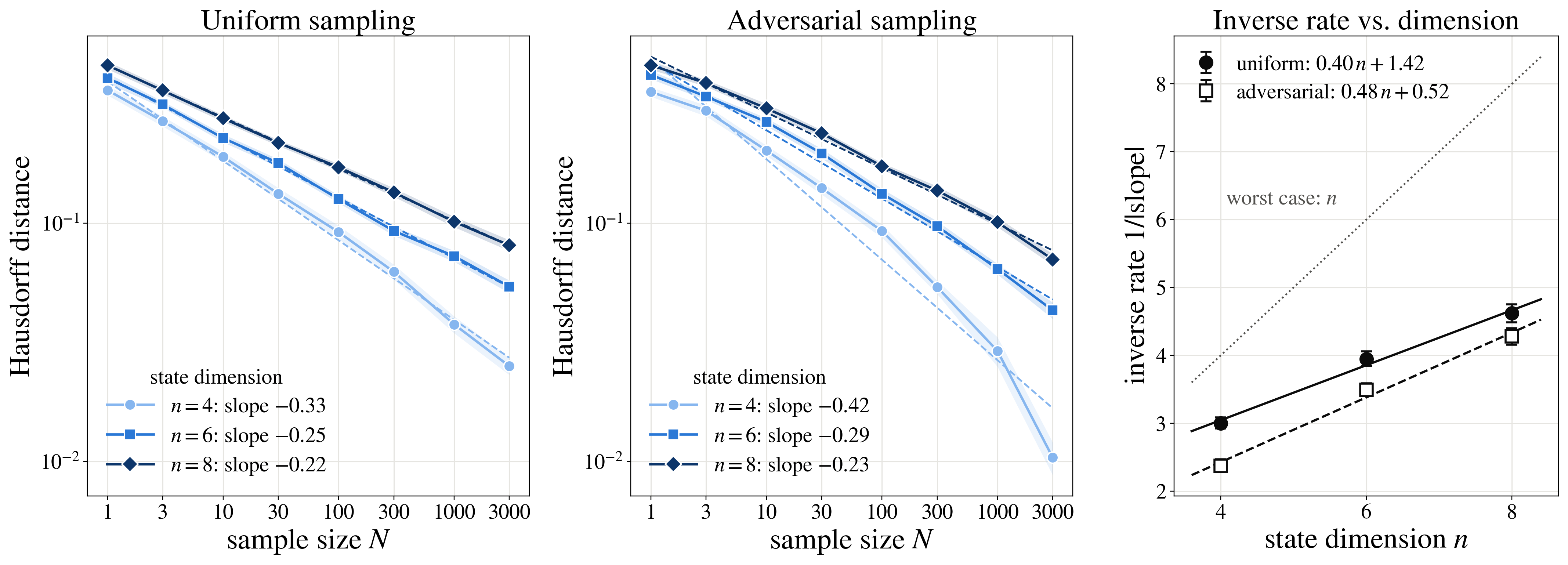}
    \caption{\latestrev{Robot arm with $n/2$ links at $T=1$. Left, middle: mean directed Hausdorff error of the convex hull of the endpoints against $N$ under uniform and adversarial sampling (bands: 95\% intervals over $30$ seeds; dashed: fitted power laws). Right: inverse rate $1/|\mathrm{slope}|$ against $n$ with its linear fit and the bootstrap intervals of Table~\ref{tab:exp-arm}; dotted: the worst-case exponent $n$.}}
    \label{fig:exp-arm}
\end{figure*}

\begin{table}[t]
\centering
\caption{Fitted convergence slopes and $95\%$ confidence intervals for the robot-arm experiments.}
\label{tab:exp-arm}
\small
\renewcommand{\arraystretch}{1.15}
\begin{tabular}{@{}ccc@{}}
\toprule
$n$ & Uniform & Adversarial\\
\midrule
$4$ & $-0.333$ & $-0.422$\\
 & {\scriptsize\latestrev{$[-0.342,-0.324]$}} & {\scriptsize\latestrev{$[-0.438,-0.405]$}}\\
$6$ & $-0.253$ & $-0.287$\\
 & {\scriptsize\latestrev{$[-0.260,-0.247]$}} & {\scriptsize\latestrev{$[-0.294,-0.279]$}}\\
$8$ & $-0.216$ & $-0.234$\\
 & {\scriptsize\latestrev{$[-0.223,-0.211]$}} & {\scriptsize\latestrev{$[-0.240,-0.227]$}}\\
\bottomrule
\end{tabular}
\end{table}

The final experiment examines dimension dependence and the empirical benefit of adaptive initial-state selection. We simulate vertical planar arms with $m\in\{2,3,4\}$ links in MuJoCo \cite{todorov2012mujoco}. The state $x=(q,v)\in\mR^n$, $n=2m$, collects joint angles and velocities; a weak inverse-dynamics controller ($K_p=0.01I$, $K_d=0.005I$) tracks a sinusoidal reference, with the remaining parameters of Liu et al.~\cite{liu2026limits}. The closed loop is time-varying. Remark~\ref{rem:lipschitz-map} extends the upper-bound argument to this setting if a Lipschitz bound for $\varphi_T$ on $S_0$ is available; we do not certify such a constant for the MuJoCo simulations. The initial set is the box $S_0=[-0.1,0.1]^n$, sampled uniformly and propagated to $T=1$; it is $(2^{-n},0.1)$-standard by Proposition~\ref{prop:standard-examples}(iii), so its geometry satisfies the upper-bound assumptions without the morphological opening used in Liu et al.~\cite{liu2026limits}. The estimator is the convex hull of the endpoints, and the error is $\max_zd(z,\conv\{Y_1,\dots,Y_N\})$ over $2000$ reference endpoints, averaged over $30$ seeds for $N\le3000$. \latestrev{Finite reference sampling and numerical projection introduce approximation error}, so the experiment shows a dimension-dependent degradation rather than a certified rate.

We compare i.i.d.\ uniform sampling with an adversarial heuristic inspired by Lew and Pavone~\cite{lew2021sampling} and approximating the update in Liu et al.~\cite[Algorithm~1]{liu2026limits}: half of the budget is drawn uniformly, and each sampled initial state is moved by a normalized step of length $0.05$ in the direction $Q(Y_j-c)$ (with zero direction left unchanged), where $c$ and $Q$ are the mean and the regularized inverse covariance of the endpoints; the moved states are projected back onto $S_0$ and propagated again. This heuristic replaces the transpose of the flow Jacobian in the novelty gradient by the identity; it does not differentiate the simulator. These samples are neither independent nor identically distributed, so Theorem~\ref{thm:lower-std} does not apply to them and the comparison is empirical.

Table~\ref{tab:exp-arm} reports slopes fitted over $N\in[1,3000]$ and \latestrev{$95\%$ confidence intervals from bootstrapping seed records while preserving shared random streams across budgets}.

Under both schemes the empirical error is well approximated by a power law whose exponent decreases in magnitude as the dimension increases (Fig.~\ref{fig:exp-arm}, Table~\ref{tab:exp-arm}). The fitted slopes are steeper than $-1/n$. This is compatible with a worst-case lower bound: the experiment uses specific systems, finite budgets, and an estimate of only one directed error. The adversarial samples also lie outside the i.i.d.\ model. With $s_n$ the fitted slope in dimension $n$ (Table~\ref{tab:exp-arm}), the inverse fitted rate $1/|s_n|$ is approximately linear over the three tested dimensions: $0.40\,n+1.42$ under uniform and $0.48\,n+0.52$ under adversarial sampling (Fig.~\ref{fig:exp-arm}, right). Adversarial sampling lowers the inverse fitted rate over this range. Its mean error is $1$ to $17\%$ larger for the tested budgets $2\le N\le100$, where half of the budget goes to perturbed copies of the same states, and better by $59\%$, $21\%$ and $13\%$ for $n=4,6,8$ at $N=3000$, a gain that shrinks with $n$.

\section{Conclusions and Open Problems}
\label{sec:conclusion}

We have characterized the number of trajectory samples needed to recover a reachable set in Hausdorff distance within the i.i.d.\ endpoint model, in terms of the local volume density $\kappa$ of $S_0$ and the one-sided Lipschitz constant $\mu$ of the vector field. Both quantities enter the upper and lower bounds through $(e^{\mu T}R/r)^n/(\rho\kappa)$; the two bounds agree up to the dimension-dependent factor $2\cdot4^n$ and logarithmic factors. For $n\ge2$, under \latestrev{the scale and domain conditions of Theorem~\ref{thm:lower-std}}, the factor $1/(\rho\kappa)$ of the upper bound is matched for every $\kappa\le\kappa_n$; the lower bound, however, is minimax over instances rather than a pointwise bound for every initial set. The planar experiments support the predicted dependence on local geometry and flow distortion. The robot-arm experiments show dimension-dependent degradation over the tested range, without certifying an asymptotic rate.

Three gaps remain. The lower bound is proved only for $\kappa\le\kappa_n$, and its prefactor $2^{-(n+1)}$ differs from the prefactor $2^n$ of the upper bound, so the two bounds do not match uniformly in the dimension. In the saturated case, i.e., when $r>2e^{\mu T}\lambda$, the upper bound \eqref{eq:upper-std-sat} no longer depends on $r$, and a matching lower bound, presumably built from initial sets with many distant pieces of size $\lambda$, is missing; the nine-piece set of Section~\ref{subsec:exp-regimes} indicates what it should look like. Finally, a single constant $\mu$ uses the largest directional expansion rate; a bound that uses the Jacobian of the flow at the corners of $S_0$ would account for the anisotropy seen in Section~\ref{subsec:exp-osl}. Beyond these gaps, the lower bound leaves adaptive and model-aware algorithms outside its scope (Remark~\ref{rem:lower-scope}).

\bibliographystyle{IEEEtran}
\bibliography{refs}

\appendices
\section{Proofs for Section~\ref{sec:regularity}}
\label{app:proofs-main}
\subsection{Flow Bounds}
\label{app:flow_bounds}

\begin{definition}[One-sided Lipschitz constant]
\label{def:osl}
A number $\mu\in\mR$ is a one-sided Lipschitz constant of $F$ on $\X$ if
\begin{align*}
    \langle F(x)-F(y),x-y\rangle\le\mu\|x-y\|^2
    \qquad\forall x,y\in\X.
\end{align*}
The constant need not be the smallest admissible one.
\end{definition}

Every $L$-Lipschitz field admits $\mu=L$, but a smaller, possibly negative, constant may be available for contracting fields \cite{lohmiller1998contraction,fb2024book}. A field need not be globally Lipschitz to admit a finite $\mu$: the cubic field $F(x)=-\|x\|^2x$ admits $\mu=0$. For $F(x)=Mx$ on $\mR^n$, the smallest admissible constant is $\mu=\lambda_{\max}(\frac12(M+M^\top))$. Thus $F(x)=-cx$ admits $\mu=-c$, the system $\dot x_1=x_1$, $\dot x_2=0$ admits $\mu=1$, and a rotation admits $\mu=0$ although $L>0$.

\begin{lemma}[Flow bounds]
\label{lem:flow-bounds}
Let Assumption~\ref{ass:well-posed} hold, let $F$ be one-sided Lipschitz with constant $\mu\in\mR$, and let $S_0$ be compact. Then
\[
    \|\varphi_T(x)-\varphi_T(y)\|\le e^{\mu T}\|x-y\|,\text{ for all }x,y\in S_0,
\]
and $S_T=\varphi_T(S_0)$ is compact.
\end{lemma}

\begin{proof}[Proof of Lemma~\ref{lem:flow-bounds}]
Fix $x,y\in S_0$ and set $w(s):=\|\varphi(s,x)-\varphi(s,y)\|$ for $s\in[0,T]$. The integral form of the flow gives
\[
    \varphi(s,x)-\varphi(s,y)
    =x-y+\int_0^s\bigl[F(\varphi(\sigma,x))-F(\varphi(\sigma,y))\bigr]\,d\sigma .
\]
Where $w(s)>0$, differentiating $w^2$ and using Definition~\ref{def:osl} gives
\begin{align*}
    w(s)w'(s)
    =& \bigl\langle \varphi(s,x)-\varphi(s,y),\,
      F(\varphi(s,x))-F(\varphi(s,y))\bigr\rangle\\
    \le& \mu\,w(s)^2,
\end{align*}
hence $w'(s)\le\mu w(s)$; by continuity the inequality holds for all $s$. Integrating over $[0,s]$ yields
\[
    w(s)\ \le\ w(0)+\mu\int_0^s w(\sigma)\,d\sigma .
\]
Gronwall's inequality \cite{fb2024book} yields
\begin{equation*}
    w(s)\ \le\ e^{\mu s}w(0),\qquad s\in[0,T].
\end{equation*}
Setting $s=T$ gives $\|\varphi_T(x)-\varphi_T(y)\|\le e^{\mu T}\|x-y\|$.
In particular $\varphi_T$ is continuous on the compact set $S_0$, so $S_T=\varphi_T(S_0)$ is compact.
\end{proof}

\begin{remark}[Lipschitz endpoint maps]
\label{rem:lipschitz-map}
Let $L_T$ be a Lipschitz constant of the endpoint map $\varphi_T$ on $S_0$; for the one-sided Lipschitz flow, $L_T=e^{\mu T}$ is admissible. The local-mass, covering, and upper-bound arguments use only this Lipschitz property. They therefore remain valid for any endpoint map with a finite $L_T$, with $L_T$ replacing $e^{\mu T}$. This includes well-posed time-varying dynamics whenever a bound on their endpoint map is available. If $L_T=0$, the image is a singleton and one sample suffices.
\end{remark}

\subsection{Endpoint Mass and Covering}
\label{app:endpoint_mass}
For reference, the local mass bound at the initial time reads as follows; Lemma~\ref{lem:endpoint-mass-std} at $T=0$ recovers it for $r\le\lambda$.

\begin{lemma}[Initial local probability lower bound]
\label{lem:initial-mass}
Let $(S_0,F,P_0)\in\Fam^{\rm std}_{R,\mu,\kappa,\lambda,\rho}$. Then, for every $x\in S_0$ and every $r\in(0,\lambda]$, $P_0(\cB_r(x))\ge\rho\,\kappa\,(r/R)^n$.
\end{lemma}

\begin{proof}
The integrated density bound and \eqref{eq:standard} give
\begin{align*}
    P_0(\cB_r(x))
    &\ \ge\ \frac{\rho}{|S_0|}\,|\cB_r(x)\cap S_0|\\
    &\ \ge\ \frac{\rho\,\kappa\,\omega_nr^n}{\omega_nR^n}
    \ =\ \rho\,\kappa\Bigl(\frac rR\Bigr)^n . \qedhere
\end{align*}
\end{proof}

Standardness is stated for radii up to $\lambda$, but it controls the local volume of $S_0$ at every scale, because a larger ball contains the ball of radius $\lambda$.

\begin{lemma}[Local volume of the initial set at every scale]
\label{lem:std-all-scales}
Let $S_0$ be $(\kappa,\lambda)$-standard. Then, for every $x\in S_0$ and every $s>0$,
\begin{equation*}
    |\cB_s(x)\cap S_0|\ \ge\ \kappa\,\omega_n\min\{s,\lambda\}^n .
\end{equation*}
\end{lemma}

\begin{proof}[Proof of Lemma~\ref{lem:std-all-scales}]
For $s\le\lambda$, the claim is Definition~\ref{def:standard}. For $s>\lambda$, the inclusion $\cB_\lambda(x)\subseteq\cB_s(x)$ and standardness at radius $\lambda$ give
\[
|\cB_s(x)\cap S_0|\ge|\cB_\lambda(x)\cap S_0|
\ge\kappa\omega_n\lambda^n.\qedhere
\]
\end{proof}

\begin{figure}[htbp]
    \centering
\begin{tikzpicture}[>=stealth, font=\small, line cap=round, line join=round, scale=0.85]
  \def\Szero{(0,0) -- (2.4,0) .. controls (3.0,1.0) and (2.3,2.4) .. (1.1,2.5) .. controls (0.2,2.55) and (-0.6,2.1) .. (-0.55,1.9) -- cycle}
  \begin{scope}
    \fill[orange!18] \Szero;
    \begin{scope}
      \clip (0,0) circle (0.8);
      \fill[blue!30] \Szero;
    \end{scope}
    \draw[orange!75!black, thick] \Szero;
    \draw[blue!65!black, dashed, thick] (0,0) circle (0.8);
    \draw[blue!65!black, ->] (0,0) -- (225:0.8) node[pos=1.32, font=\scriptsize] {$s$};
    \fill (0,0) circle (1.4pt);
    \node[below right, inner sep=1.5pt] at (0,0) {$x$};
    \node at (1.55,1.45) {$S_0$};
    \node[blue!65!black, font=\scriptsize] at (1.05,-1.05) {$\cB_s(x)\cap S_0$};
  \end{scope}
  \draw[->, thick] (3.15,1.25) -- (4.25,1.25) node[midway, above, inner sep=2pt] {$\varphi_T$};
  \begin{scope}[xshift=5.1cm]
    \begin{scope}[cm={1.25,0.35,0.15,0.85,(0,0)}]
      \fill[orange!18] \Szero;
      \begin{scope}
        \clip (0,0) circle (0.8);
        \fill[blue!30] \Szero;
      \end{scope}
      \draw[orange!75!black, thick] \Szero;
      \draw[blue!65!black, dashed, thick] (0,0) circle (0.8);
    \end{scope}
    \draw[gray!55!black, thin] (0,0) circle (1.16);
    \draw[gray!55!black, ->] (0,0) -- (250:1.16) node[pos=1.22, font=\scriptsize] {$r$};
    \fill (0,0) circle (1.4pt);
    \node[below right, inner sep=1.5pt] at (0,0) {$y$};
    \node at (2.45,2.0) {$S_T$};
    \node[gray!55!black, font=\scriptsize] at (-1.6,0.35) {$\cB_r(y)$};
    \node[blue!65!black, font=\scriptsize] at (1.15,-1.5) {$\varphi_T(\cB_s(x)\cap S_0)$};
  \end{scope}
\end{tikzpicture}
    \caption{Lemma~\ref{lem:endpoint-mass-std}: $\varphi_T$ maps $\cB_s(x)\cap S_0$ (shaded, left), $s=e^{-\mu T}r$, into $\cB_r(y)\cap S_T$ (right), $y=\varphi_T(x)$. Dashed: the boundary of $\cB_s(x)$ and its image; thin: $\cB_r(y)$.}
    \label{fig:proof-ball-inclusion}
\end{figure}
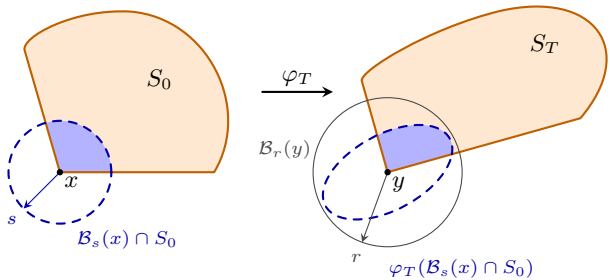

\begin{proof}[Proof of Lemma~\ref{lem:endpoint-mass-std}]
Fix $y\in S_T$ and choose $x\in S_0$ with $y=\varphi_T(x)$. Set $s=e^{-\mu T}r$. By Lemma~\ref{lem:flow-bounds}, every $z\in\cB_s(x)\cap S_0$ satisfies
\[
\|\varphi_T(z)-y\|\le e^{\mu T}\|z-x\|\le r.
\]
Thus $\cB_s(x)\cap S_0\subseteq\varphi_T^{-1}(\cB_r(y)\cap S_T)$; see Fig.~\ref{fig:proof-ball-inclusion}. The pushforward identity, the integrated density bound, and Lemma~\ref{lem:std-all-scales} yield
\begin{align*}
P_T(\cB_r(y)\cap S_T)
&\ge P_0(\cB_s(x)\cap S_0)\\
&\ge\frac{\rho}{\omega_nR^n}|\cB_s(x)\cap S_0|\\
&\ge\rho\kappa\left(\frac{\min\{s,\lambda\}}{R}\right)^n.
\end{align*}
Substituting $s=e^{-\mu T}r$ proves \eqref{eq:GammaT-std}.
\end{proof}

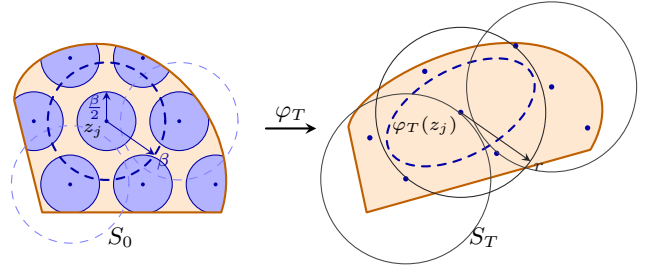
\begin{figure}[htbp]
    \centering
\begin{tikzpicture}[>=stealth, font=\small, line cap=round, line join=round, scale=0.74]
  \def\Szero{(0,0) -- (3.2,0) .. controls (3.6,1.2) and (2.6,2.8) .. (1.2,3.0) .. controls (0.3,3.1) and (-0.5,2.6) .. (-0.5,2.0) -- cycle}
  \def\pts{0.5/0.5,1.8/0.5,3.1/0.5,-0.15/1.63,1.15/1.63,2.45/1.63,0.5/2.76,1.8/2.76}
  \begin{scope}
    \fill[orange!18] \Szero;
    \begin{scope}
      \clip \Szero;
      \foreach \x/\y in \pts {
        \fill[blue!30] (\x,\y) circle (0.525);
        \draw[blue!65!black, thin] (\x,\y) circle (0.525);
      }
    \end{scope}
    \draw[orange!75!black, thick] \Szero;
    \foreach \x/\y in \pts { \fill[blue!65!black] (\x,\y) circle (1pt); }
    \foreach \x/\y in {0.5/0.5,2.45/1.63} { \draw[blue!50, dashed, thin] (\x,\y) circle (1.05); }
    \draw[blue!65!black, dashed, thick] (1.15,1.63) circle (1.05);
    \draw[blue!65!black, ->] (1.15,1.63) -- (1.15,2.155) node[pos=0.5, left, inner sep=1pt, font=\scriptsize] {$\tfrac{\beta}{2}$};
    \draw[blue!65!black, ->] (1.15,1.63) -- ++(-35:1.05) node[pos=1.18, font=\scriptsize] {$\beta$};
    \node[below left, inner sep=1pt, font=\scriptsize] at (1.15,1.63) {$z_j$};
    \node at (1.4,-0.45) {$S_0$};
  \end{scope}
  \draw[->, thick] (4.0,1.5) -- (4.9,1.5) node[midway, above, inner sep=2pt] {$\varphi_T$};
  \begin{scope}[xshift=5.8cm]
    \begin{scope}[cm={1.25,0.35,0.15,0.85,(0,0)}]
      \fill[orange!18] \Szero;
      \draw[orange!75!black, thick] \Szero;
      \foreach \x/\y in \pts { \node[circle, fill=blue!65!black, inner sep=0pt, minimum size=2pt] at (\x,\y) {}; }
      \draw[blue!65!black, dashed, thick] (1.15,1.63) circle (1.05);
      \coordinate (zTj) at (1.15,1.63);
      \coordinate (zTa) at (0.5,0.5);
      \coordinate (zTb) at (2.45,1.63);
    \end{scope}
    \draw[gray!55!black, thin] (zTa) circle (1.52);
    \draw[gray!55!black, thin] (zTb) circle (1.52);
    \draw[gray!30!black, thin] (zTj) circle (1.52);
    \draw[gray!30!black, ->] (zTj) -- ++(-35:1.52) node[pos=1.12, font=\scriptsize] {$r$};
    \node[below left, inner sep=1pt, font=\scriptsize] at (zTj) {$\varphi_T(z_j)$};
    \node at (2.1,-0.45) {$S_T$};
  \end{scope}
\end{tikzpicture}
    \caption{Lemma~\ref{lem:covering-std}. Left: a maximal $\beta$-separated set $\{z_j\}$ in $S_0$; the disjoint shaded sets $\cB_{\beta/2}(z_j)\cap S_0$ bound $M$, and the dashed balls $\cB_\beta(z_j)$ cover $S_0$. Right: the images $\varphi_T(z_j)$ form an $r$-cover of $S_T$, $r=e^{\mu T}\beta$; dashed: the image of $\cB_\beta(z_j)$; thin: the balls of radius $r$.}
    \label{fig:proof-covering}
\end{figure}

\begin{proof}[Proof of Lemma~\ref{lem:covering-std}]
Fix $r>0$ and set $\beta=e^{-\mu T}r$. The image under $\varphi_T$ of any $\beta$-net in $S_0$ is an $r$-net in $S_T$, because $\varphi_T$ is $e^{\mu T}$-Lipschitz. Hence
\begin{equation}
\cN(S_T,r)\le\cN(S_0,\beta).
\label{eq:covering-reduction}
\end{equation}
Choose a maximal $\beta$-separated subset $\{z_1,\dots,z_M\}$ of $S_0$, where separation means $\|z_i-z_j\|>\beta$ for $i\ne j$. Compactness ensures that this set is finite. Maximality implies that it is a $\beta$-net: otherwise another point could be added. The balls $\cB_{\beta/2}(z_j)$ are pairwise disjoint, so Lemma~\ref{lem:std-all-scales} gives
\begin{align*}
\omega_nR^n=|S_0|
&\ge\sum_{j=1}^M|\cB_{\beta/2}(z_j)\cap S_0|\\
&\ge M\kappa\omega_n\min\{\beta/2,\lambda\}^n.
\end{align*}
Consequently,
\begin{equation*}
\cN(S_0,\beta)\le M\le
\frac{R^n}{\kappa\min\{\beta/2,\lambda\}^n}.
\end{equation*}
Combining this with \eqref{eq:covering-reduction} proves \eqref{eq:covering-std}. Figure~\ref{fig:proof-covering} illustrates the packing and covering steps.
\end{proof}

\subsection{Upper Bound}
\label{app:upper_bound}
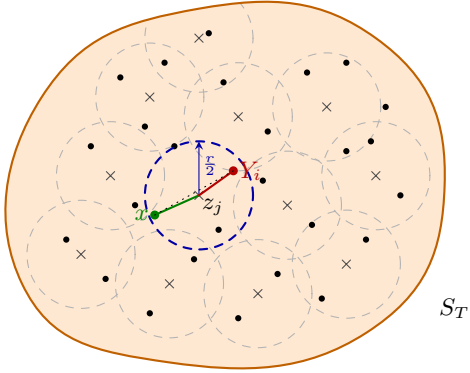
\begin{figure}[htbp]
    \centering
\begin{tikzpicture}[>=stealth, font=\small, line cap=round, line join=round, scale=1.3]
  \def\STblob{plot[smooth cycle, tension=0.8] coordinates {(0,0.5) (1.5,-0.3) (3.5,0) (4.3,1.5) (3.6,3.0) (1.8,3.3) (0.2,2.4)}}
  \fill[orange!18] \STblob;
  \begin{scope}
    \clip \STblob;
    \foreach \x/\y in {0.6/0.8,1.5/0.5,2.4/0.4,3.3/0.7,0.9/1.6,2.7/1.3,3.6/1.6,1.3/2.4,2.2/2.2,3.1/2.3,1.8/3.0} {
      \draw[gray!60, dashed, thin] (\x,\y) circle (0.55);
    }
  \end{scope}
  \draw[orange!75!black, thick] \STblob;
  \foreach \x/\y in {0.6/0.8,1.5/0.5,2.4/0.4,3.3/0.7,0.9/1.6,2.7/1.3,3.6/1.6,1.3/2.4,2.2/2.2,3.1/2.3,1.8/3.0} {
    \draw[gray!60!black] (\x,\y) ++(-1.2pt,-1.2pt) -- ++(2.4pt,2.4pt) (\x,\y) ++(-1.2pt,1.2pt) -- ++(2.4pt,-2.4pt);
  }
  \foreach \x/\y in {0.45/0.95,0.85/0.55,1.3/0.2,1.75/0.75,2.2/0.15,2.6/0.6,3.05/0.35,3.5/0.95,3.85/1.4,0.7/1.9,1.15/1.3,1.55/1.9,2.0/1.05,2.45/1.55,2.95/1.1,3.3/1.85,3.7/2.3,1.0/2.6,1.45/2.75,2.05/2.55,2.5/2.05,2.9/2.65,3.3/2.75,1.9/3.05,1.25/2.1,3.55/1.95} {
    \fill[black] (\x,\y) circle (0.9pt);
  }
  \draw[blue!65!black, dashed, thick] (1.8,1.4) circle (0.55);
  \draw[blue!65!black, ->] (1.8,1.4) -- (1.8,1.95) node[pos=0.55, right, inner sep=1pt, font=\scriptsize] {$\tfrac r2$};
  \draw[gray!60!black] (1.8,1.4) ++(-1.2pt,-1.2pt) -- ++(2.4pt,2.4pt) (1.8,1.4) ++(-1.2pt,1.2pt) -- ++(2.4pt,-2.4pt);
  \node[below right, inner sep=1pt] at (1.8,1.4) {$z_j$};
  \fill[red!70!black] (2.15,1.65) circle (1.3pt);
  \node[right, inner sep=2pt, red!70!black] at (2.15,1.65) {$Y_i$};
  \fill[green!50!black] (1.35,1.2) circle (1.3pt);
  \node[left, inner sep=2pt, green!50!black] at (1.35,1.2) {$x$};
  \draw[green!50!black, thick] (1.35,1.2) -- (1.8,1.4);
  \draw[red!70!black, thick] (1.8,1.4) -- (2.15,1.65);
  \draw[black, dotted] (1.35,1.2) -- (2.15,1.65);
  \node[anchor=west] at (4.15,0.25) {$S_T$};
\end{tikzpicture}
    \caption{Theorem~\ref{thm:upper-std}: the centers $z_j$ (crosses) of an $\frac r2$-net of $S_T$ with their balls, and the endpoint samples (dots). When every net ball contains a sample, $d(x,\widehat S_N)\le\|x-z_j\|+\|z_j-Y_i\|\le r$ for every $x\in S_T$.}
    \label{fig:proof-net}
\end{figure}

\begin{proof}[Proof of Theorem~\ref{thm:upper-std}]
For cases (1) and (2), write $\varrho=\varrho_T(r)$ and $q_r=\rho\kappa(\varrho/R)^n$. By Lemma~\ref{lem:endpoint-mass-std}, every ball of radius $r/2$ centered in $S_T$ has probability at least $q_r$. Choose an $r/2$-net $\{z_1,\dots,z_M\}\subseteq S_T$ of minimum cardinality. Lemma~\ref{lem:covering-std} and the inequality
\[
\min\{e^{-\mu T}r/4,\lambda\}
\ge\tfrac12\min\{e^{-\mu T}r/2,\lambda\}=\varrho/2
\]
give
\begin{equation}
M\le\frac{2^nR^n}{\kappa\varrho^n}.
\label{eq:upper-net-size}
\end{equation}
Let $E_j$ be the event that $\cB_{r/2}(z_j)$ contains no endpoint. Independence and the union bound imply
\begin{align*}
P_T^N\Bigl(\bigcup_{j=1}^M E_j\Bigr)
&\le M(1-q_r)^N\le Me^{-Nq_r}.
\end{align*}
Both sample-size conditions are obtained from
\[
N\ge\frac1{q_r}
\left[\log\frac{2^nR^n}{\kappa\varrho^n}+\log\frac1\delta\right]
\]
by substituting the relevant value of $\varrho$. Under either condition, \eqref{eq:upper-net-size} gives $Me^{-Nq_r}\le\delta$.

Outside the union of the bad events, every $x\in S_T$ has a net center $z_j$ and an endpoint $Y_i$ such that $\|x-z_j\|\le r/2$ and $\|z_j-Y_i\|\le r/2$. Since $Y_i\in\widehat S_N$,
\[
d(x,\widehat S_N)\le\|x-Y_i\|
\le\|x-z_j\|+\|z_j-Y_i\|\le r;
\]
see Fig.~\ref{fig:proof-net}.

For case (3), $\diam(S_T)\le e^{\mu T}D_0\le r$. Almost surely, $Y_1\in S_T\cap\widehat S_N$, so $d(x,\widehat S_N)\le\|x-Y_1\|\le r$ for every $x\in S_T$.
\end{proof}

\begin{proof}[Proof of Corollary~\ref{cor:hausdorff-outer-deviation}]
Theorem~\ref{thm:upper-std} provides an event of probability at least $1-\delta$ on which $\sup_{x\in S_T}d(x,\widehat S_N)\le r$. The outer-deviation condition $\widehat S_N\subseteq\cB_\eta(S_T)$ means that $d(z,S_T)\le\eta$ for every $z\in\widehat S_N$. Therefore, on the same event,
\begin{align*}
    d_H(S_T,\widehat S_N)
    &\ =\ \max\Bigl\{\sup_{x\in S_T}d(x,\widehat S_N),\ \sup_{z\in\widehat S_N}d(z,S_T)\Bigr\}\\
    &\ \le\ \max\{r,\eta\}. \qedhere
\end{align*}
\end{proof}

\begin{proof}[Proof of Corollary~\ref{cor:convex-upper}]
First, $R\le D$: fix $x_0\in S_0$. Every point of $S_0$ is within distance $D$ of $x_0$, so $S_0\subseteq\cB_D(x_0)$, and $\omega_nR^n=|S_0|\le\omega_nD^n$. Hence $(R/D)^n\in(0,1]$.

Next, every ball centered in $S_0$ contains a scaled copy of $S_0$. Fix $x\in S_0$ and $s\in(0,D]$, and consider the homothety
\begin{align*}
    H_x(z):=x+\frac sD\,(z-x),\qquad z\in\mR^n .
\end{align*}
For $z\in S_0$, convexity of $S_0$ gives $H_x(z)=(1-\frac sD)x+\frac sDz\in S_0$, and
\begin{align*}
    \|H_x(z)-x\|\ =\ \frac sD\,\|z-x\|\ \le\ \frac sD\,D\ =\ s .
\end{align*}
Hence $H_x(S_0)\subseteq\cB_s(x)\cap S_0$. Since $H_x$ multiplies volumes by $(s/D)^n$ and $|S_0|=\omega_nR^n$,
\begin{align*}
    |\cB_s(x)\cap S_0|
    &\ \ge\ |H_x(S_0)|
    \ =\ \Bigl(\frac sD\Bigr)^n|S_0|\\
    &\ =\ \Bigl(\frac RD\Bigr)^n\omega_ns^n .
\end{align*}
This is $((R/D)^n,D)$-standardness.

By this standardness, the instance belongs to $\Fam^{\rm std}_{R,\mu,\kappa',\lambda',\rho}$ with $(\kappa',\lambda'):=((R/D)^n,D)$, so Theorem~\ref{thm:upper-std} applies with $(\kappa,\lambda)$ replaced by $(\kappa',\lambda')$, that is, with $\varrho_T(r)$ replaced by $\varrho'_T(r)$. For this pair,
\begin{align*}
    \frac1{\rho\,\kappa'}\Bigl(\frac{R}{\varrho'_T(r)}\Bigr)^n
    &\ =\ \frac1\rho\Bigl(\frac{D}{\varrho'_T(r)}\Bigr)^n,\\
    \log\Bigl(\frac{2^nR^n}{\kappa'\,\varrho'_T(r)^n}\Bigr)
    &\ =\ n\log\frac{2D}{\varrho'_T(r)},
\end{align*}
so the bound of Theorem~\ref{thm:upper-std} becomes \eqref{eq:upper-convex}.
\end{proof}

\subsection{Standard Sets and Inner Cones}
\label{app:geometry}

This subsection collects the cone and inner-cone facts used in the lower-bound construction of Appendix~\ref{app:lower_bound}.

A set $K\subseteq\mR^n$ is a \emph{cone} if $tK\subseteq K$ for all $t>0$. For a closed cone $K$, the \emph{normalized solid angle} $\Theta(K):=|K\cap\cB_1|/\omega_n\in[0,1]$ is the fraction of every ball centered at the apex that $K$ occupies, because $|K\cap\cB_r|=\Theta(K)\,\omega_nr^n$ by homogeneity; a planar wedge of opening $\theta$ has $\Theta=\theta/(2\pi)$. For $u\in\mathbb S^{n-1}$ and $\alpha\in(0,\pi/2]$, $K_\alpha(u):=\{v:\langle v,u\rangle\ge\|v\|\cos\alpha\}$ is the circular cone with axis $u$ and half-angle $\alpha$, and $\vartheta_n(\alpha):=\Theta(K_\alpha(u))$, which does not depend on $u$; in particular, the constant of \eqref{eq:kappa-n} is $\kappa_n=\vartheta_n(\pi/4)$.

\begin{definition}[Uniform inner-cone condition]
\label{def:uic}
Let $\cK$ be a family of closed convex cones with nonempty interior and let $h>0$. For $x\in S_0$ let $\cK_h(x):=\{K\in\cK:x+(K\cap\cB_h)\subseteq S_0\}$ be the \emph{admissible} cones at $x$. The set $S_0$ satisfies $\UIC(\cK,h)$ if $\cK_h(x)\ne\varnothing$ for every $x\in S_0$. The \emph{aperture} of $\cK$ is $\vartheta(\cK):=\inf_{K\in\cK}\Theta(K)$. For the family $\cK=\{K_\alpha(u)\}_{u\in\mathbb S^{n-1}}$ of circular cones of half-angle $\alpha$ we write $\UIC(\alpha,h)$; its aperture is $\vartheta_n(\alpha)$.
\end{definition}

\paragraph{Inner cones} The next two lemmas combine the solid angle of a circular cone with the volume that an inner cone guarantees.

\begin{lemma}[Solid angle of circular cones]
\label{lem:circular-solid-angle}
For $n\ge2$ and $\alpha\in(0,\pi/2]$,
\begin{equation}
    \vartheta_n(\alpha)=\frac{(n-1)\,\omega_{n-1}}{n\,\omega_n}\int_0^\alpha\sin^{n-2}\phi\,d\phi
    \ \ge\ \Bigl(\frac{\sin\alpha}{1+\sin\alpha}\Bigr)^n,
    \label{eq:vartheta-formula}
\end{equation}
with $\vartheta_2(\alpha)=\alpha/\pi$ and $\vartheta_3(\alpha)=(1-\cos\alpha)/2$. At $\alpha=\pi/2$ the cone $K_{\pi/2}(u)$ is the half-space $\{v:\langle v,u\rangle\ge0\}$, so $\vartheta_n(\pi/2)=\frac12$, while the lower bound in \eqref{eq:vartheta-formula} equals $2^{-n}$, the positive-reach constant.
\end{lemma}

\begin{proof}
\emph{Formula.} Write $\xi=t\theta$ with $t=\|\xi\|\in[0,1]$ and $\theta\in\mathbb S^{n-1}$, and $\theta=\cos\phi\,u+\sin\phi\,\eta$ with the polar angle $\phi=\angle(\theta,u)\in[0,\pi]$ and $\eta$ in the unit sphere of $u^\perp$, a copy of $\mathbb S^{n-2}$. Then \rev{$d\xi=t^{n-1}\,dt\,d\sigma_{n-1}(\theta)$ and $d\sigma_{n-1}(\theta)=\sin^{n-2}\phi\,d\phi\,d\sigma_{n-2}(\eta)$}, and $\theta\in K_\alpha(u)$ if and only if $\phi\le\alpha$. Hence, with \rev{$\sigma_{n-2}(\mathbb S^{n-2})=(n-1)\omega_{n-1}$},
\begin{align*}
    |K_\alpha(u)\cap\cB_1|
    &=\int_{\mathbb S^{n-1}\cap K_\alpha(u)}\int_0^1t^{n-1}\,dt\,d\rev{\sigma_{n-1}}(\theta)\\
    &=\frac1n\,\rev{\sigma_{n-1}}\bigl(\mathbb S^{n-1}\cap K_\alpha(u)\bigr)\\
    &=\frac1n\,\rev{\sigma_{n-2}(\mathbb S^{n-2})}\int_0^\alpha\sin^{n-2}\phi\,d\phi\\
    &=\frac{(n-1)\,\omega_{n-1}}{n}\int_0^\alpha\sin^{n-2}\phi\,d\phi ,
\end{align*}
and dividing by $\omega_n$ gives the formula. With $\omega_1=2$, $\omega_2=\pi$ and $\omega_3=4\pi/3$,
\begin{align*}
    \vartheta_2(\alpha)&=\frac{2}{2\pi}\int_0^\alpha d\phi=\frac\alpha\pi,\\
    \vartheta_3(\alpha)&=\frac{2\pi}{4\pi}\int_0^\alpha\sin\phi\,d\phi=\frac{1-\cos\alpha}{2}.
\end{align*}

\emph{Lower bound.} We fit a ball into $K_\alpha(u)\cap\cB_1$. A ball centered at $tu$, $t>0$, lies in the cone if its radius is at most $t\sin\alpha$, the distance from $tu$ to the boundary of the cone, and in $\cB_1$ if, in addition, $t+t\sin\alpha\le1$. The largest such ball has $t=1/(1+s)$, where $s:=\sin\alpha$; that is, it has center and radius
\begin{align*}
    c:=\frac{u}{1+s},\qquad
    \varrho:=\frac{s}{1+s}=s\,\|c\|.
\end{align*}
Let $z=c+w$ with $\|w\|\le\varrho$. By the triangle inequality, $\|z\|\le\|c\|+\|w\|\le\frac1{1+s}+\frac{s}{1+s}=1$. To check the cone condition, split $w=au+w_\perp$ with $a:=\langle w,u\rangle$ and $w_\perp\perp u$, and let $b:=\|w_\perp\|$, so that $a^2+b^2=\|w\|^2$. Then $z=(\|c\|+a)u+w_\perp$, so
\begin{align*}
    \langle z,u\rangle=\|c\|+a,\qquad
    \|z\|^2=(\|c\|+a)^2+b^2 .
\end{align*}
By definition, $z\in K_\alpha(u)$ means $\langle z,u\rangle\ge\|z\|\cos\alpha$. This holds as soon as
\begin{align}
    b\cos\alpha\ \le\ (\|c\|+a)\sin\alpha .
    \label{eq:cone-test}
\end{align}
Indeed, \eqref{eq:cone-test} and $\sin\alpha>0$ give $\|c\|+a\ge0$, and squaring \eqref{eq:cone-test},
\begin{align*}
    (\|c\|+a)^2
    &=(\|c\|+a)^2\cos^2\alpha+(\|c\|+a)^2\sin^2\alpha\\
    &\ge(\|c\|+a)^2\cos^2\alpha+b^2\cos^2\alpha\ =\ \|z\|^2\cos^2\alpha ,
\end{align*}
so $\langle z,u\rangle=\|c\|+a\ge\|z\|\cos\alpha$.

It remains to prove \eqref{eq:cone-test}. By the Cauchy--Schwarz inequality for the vectors $(b,-a)$ and $(\cos\alpha,\sin\alpha)$ of $\mR^2$, and by $\varrho=\|c\|\sin\alpha$,
\begin{align*}
    b\cos\alpha-a\sin\alpha
    &=\bigl\langle(b,-a),(\cos\alpha,\sin\alpha)\bigr\rangle\\
    &\le\sqrt{a^2+b^2}\,\sqrt{\cos^2\alpha+\sin^2\alpha}\\
    &=\|w\|\ \le\ \varrho\ =\ \|c\|\sin\alpha ,
\end{align*}
which is \eqref{eq:cone-test} after moving $a\sin\alpha$ to the right.

Hence $\cB_\varrho(c)\subseteq K_\alpha(u)\cap\cB_1$, and
\begin{align*}
    \vartheta_n(\alpha)=\frac{|K_\alpha(u)\cap\cB_1|}{\omega_n}\ \ge\ \frac{|\cB_\varrho(c)|}{\omega_n}=\varrho^n=\Bigl(\frac{\sin\alpha}{1+\sin\alpha}\Bigr)^n . \qedhere
\end{align*}
\end{proof}

\begin{lemma}[Inner cones imply standardness]
\label{lem:uic-standard}
If $S_0$ is compact, satisfies $\UIC(\cK,h)$, and $\vartheta(\cK)>0$, then $S_0$ is $(\vartheta(\cK),h)$-standard.
\end{lemma}

\begin{proof}
For $x\in S_0$, $r\le h$ and $K\in\cK_h(x)$, $x+(K\cap\cB_r)\subseteq\cB_r(x)\cap S_0$, so
\begin{align*}
    |\cB_r(x)\cap S_0|
    &\ \ge\ |K\cap\cB_r|
    \ =\ \Theta(K)\,\omega_nr^n\\
    &\ \ge\ \vartheta(\cK)\,\omega_nr^n . \qedhere
\end{align*}
\end{proof}

\begin{proposition}[A ball satisfies the inner-cone condition]
\label{prop:reach-implies-uic}
A closed ball $\cB_{r_0}(c)$ satisfies $\UIC(\alpha,r_0\cos\alpha)$ for every $\alpha\in(0,\pi/2)$.
\end{proposition}

\begin{proof}
Fix $x\in\cB_{r_0}(c)$, let $d:=\|x-c\|$ and $u:=(c-x)/d$ (any unit vector if $d=0$), and let $v$ be a unit vector with $\phi:=\angle(v,u)\le\alpha$. Let $\tau:=r_0\cos\alpha$. The function $f(t):=\|x+tv-c\|^2=t^2-2td\cos\phi+d^2$ is convex with $f(0)=d^2\le r_0^2$. Since $g(d):=d^2-2\tau d\cos\alpha+\tau^2$ is convex on $[0,r_0]$,
\begin{align*}
    f(\tau)&\le g(d)\le\max\{g(0),g(r_0)\}\\
    &=\max\{r_0^2\cos^2\alpha,\ r_0^2\sin^2\alpha\}\le r_0^2 .
\end{align*}
Hence $x+tv\in\cB_{r_0}(c)$ for $t\in[0,\tau]$, i.e.\ $x+(K_\alpha(u)\cap\cB_\tau)\subseteq\cB_{r_0}(c)$.
\end{proof}

\subsection{Lower Bound}
\label{app:lower_bound}

Throughout this subsection, $n\ge2$, $y=(y_1,y')\in\mR\times\mR^{n-1}$, $\alpha\in(0,\pi/4]$ and $K:=K_\alpha(-e_1)=\{w:\ w_1\le0,\ \|w'\|\le-w_1\tan\alpha\}$, so that $\Theta(K)=\vartheta_n(\alpha)$.

\begin{lemma}[Cutting a tip]
\label{lem:cut-tip}
Let $q\in\mR^n$ and $\varsigma>0$. If $z\in q+K$ and $\|z-q\|\ge\varsigma$, then $z+K\subseteq(q+K)\setminus\mathring{\cB}_\varsigma(q)$, where $\mathring{\cB}_\varsigma(q)$ is the open ball.
\end{lemma}

\begin{proof}
As $K$ is a convex cone, $z+K\subseteq q+K$. Two vectors of $K$ make an angle of at most $2\alpha\le\pi/2$, so $\langle z-q,w\rangle\ge0$ for $w\in K$, and
\begin{align*}
    \|z+w-q\|^2
    &\ =\ \|z-q\|^2+2\langle z-q,w\rangle+\|w\|^2\\
    &\ \ge\ \|z-q\|^2\ \ge\ \varsigma^2 . \qedhere
\end{align*}
\end{proof}

\begin{figure}[htbp]
    \centering
\begin{tikzpicture}[>=stealth, font=\small, line cap=round, line join=round, scale=0.85]
  \begin{scope}
    \fill[orange!18] (0,0) circle (1.3);
    \fill[orange!18] (1.56,0) -- (0,0.702) -- (0,-0.702) -- cycle;
    \fill[orange!18] (-1.56,0) -- (0,0.702) -- (0,-0.702) -- cycle;
    \fill[blue!35] (1.56,0) -- ++(155.77:0.19) arc[start angle=155.77, end angle=204.23, radius=0.19] -- cycle;
    \fill[white] (-1.56,0) -- ++(-24.23:0.19) arc[start angle=-24.23, end angle=24.23, radius=0.19] -- cycle;
    \fill[pattern=north east lines, pattern color=blue!65!black]
      (-1.56,0) -- ++(-24.23:0.19) arc[start angle=-24.23, end angle=24.23, radius=0.19] -- cycle;
    \begin{scope}[even odd rule]
      \clip (0,0) circle (1.3) (-2.2,-2.2) rectangle (2.2,2.2);
      \draw[orange!75!black, thick] (1.56,0) -- (0,0.702) (1.56,0) -- (0,-0.702);
      \draw[orange!75!black, thick] (-1.56,0) -- (0,0.702) (-1.56,0) -- (0,-0.702);
    \end{scope}
    \draw[orange!75!black, thick] (0,0) circle (1.3);
    \draw[blue!65!black, dashed] (-1.56,0) circle (0.19);
    \fill (1.56,0) circle (1.2pt);
    \fill (-1.56,0) circle (1.2pt);
    \node[below, inner sep=3pt, font=\scriptsize] at (1.56,-0.05) {$q_0$};
    \node[below, inner sep=3pt, font=\scriptsize] at (-1.56,-0.19) {$q_1$};
    \draw[->, thin] (-1.56,0) -- ++(-150:0.19) node[pos=1, left, inner sep=1pt, font=\scriptsize] {$\varsigma$};
    \fill (0,0) circle (1pt);
    \draw[->, thin] (0,0) -- (120:1.3) node[pos=0.55, right, inner sep=1pt, font=\scriptsize] {$\bar R$};
    \node[blue!65!black, font=\scriptsize] at (1.56,0.42) {$W_0$};
    \node[blue!65!black, font=\scriptsize] at (-1.56,0.6) {$W_1$};
    \node at (0,-1.9) {thin tips, $\tan\alpha\le\frac12$};
  \end{scope}
  \begin{scope}[xshift=5.3cm]
    \def\hull{(1.711,0) -- (50:1.1) arc[start angle=50, end angle=130, radius=1.1] -- (-1.711,0) -- (230:1.1) arc[start angle=230, end angle=310, radius=1.1] -- cycle}
    \fill[orange!18] \hull;
    \fill[blue!35] (1.711,0) -- ++(140:0.32) arc[start angle=140, end angle=220, radius=0.32] -- cycle;
    \fill[white] (-1.711,0) -- ++(-40:0.32) arc[start angle=-40, end angle=40, radius=0.32] -- cycle;
    \fill[pattern=north east lines, pattern color=blue!65!black]
      (-1.711,0) -- ++(-40:0.32) arc[start angle=-40, end angle=40, radius=0.32] -- cycle;
    \draw[orange!75!black, thick] \hull;
    \draw[orange!75!black, thin, dotted] (0,0) circle (1.1);
    \draw[blue!65!black, dashed] (-1.711,0) circle (0.32);
    \fill (1.711,0) circle (1.2pt);
    \fill (-1.711,0) circle (1.2pt);
    \node[below, inner sep=3pt, font=\scriptsize] at (1.711,-0.05) {$q_0$};
    \node[below, inner sep=3pt, font=\scriptsize] at (-1.711,-0.32) {$q_1$};
    \draw[thin] (1.711,0) ++(140:0.62) arc[start angle=140, end angle=180, radius=0.62];
    \node[font=\scriptsize] at (1.0,0.13) {$\alpha$};
    \fill (0,0) circle (1pt);
    \draw[->, thin] (0,0) -- (100:1.1) node[pos=0.55, right, inner sep=1pt, font=\scriptsize] {$\bar R$};
    \node[blue!65!black, font=\scriptsize] at (1.711,0.4) {$W_0$};
    \node[blue!65!black, font=\scriptsize] at (-1.711,0.55) {$W_1$};
    \node at (0,-1.9) {wide tips, $\tan\alpha>\frac12$};
  \end{scope}
\end{tikzpicture}
    \caption{Hard instances of Theorem~\ref{thm:lower-std} for thin tips (left) and wide tips (right). $S^{(0)}$ keeps the tip at $q_0$ and cuts off the tip at $q_1$ by $\mathring{\cB}_\varsigma(q_1)$ (dashed); $S^{(1)}$ does the reverse. The instances differ on the sectors $W_0$ (shaded) and $W_1$ (hatched).}
    \label{fig:proof-lower-instances}
\end{figure}
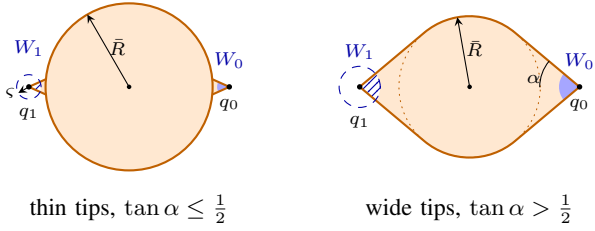

The hard instances share a body $\Omega$ with conical tips at $\pm q$, $q:=ae_1$, where $\Omega$ and $a$ depend on $\bar R>0$ and on the width of the tips. For thin tips, $\tan\alpha\le\frac12$, let $a:=\frac65\bar R$ and
\begin{align*}
    \Lambda&:=(q+K)\cap\{y_1\ge0\},\\
    \Omega&:=\cB_{\bar R}\cup\Lambda\cup(-\Lambda),
\end{align*}
a ball with two conical spikes. For wide tips, $\frac12<\tan\alpha\le1$, let $a:=\bar R/\sin\alpha$, $b:=\bar R\sin\alpha$ and
\begin{align*}
    \Lambda&:=(q+K)\cap\{y_1\ge b\},\\
    \Omega&:=\conv\bigl(\cB_{\bar R}\cup\{q,-q\}\bigr).
\end{align*}
In both cases $\Omega$ scales with $\bar R$. With $q_0:=q$ and $q_1:=-q$, the instances and their tip sectors are
\[
    S^{(i)}:=\Omega\setminus\mathring{\cB}_\varsigma(q_{1-i}),\qquad
    W_i:=\Omega\cap\cB_\varsigma(q_i),\qquad i=0,1,
\]
so $S^{(i)}$ keeps the tip at $q_i$ and cuts off the other one (Fig.~\ref{fig:proof-lower-instances}).

\begin{lemma}[Shape of the body]
\label{lem:tip-body}
(a) For thin tips, $\Omega\subseteq\cB_{6\bar R/5}$ and $(q+K)\cap\{-\bar R/2\le y_1\le0\}\subseteq\cB_{\bar R}$.
(b) For wide tips, $\Omega\subseteq q+K$, $\Omega=\cB_{\bar R}\cup\Lambda\cup(-\Lambda)$ and $|\Omega|\le\frac{13}{5}\,\omega_n\bar R^n$.
\end{lemma}

\begin{proof}
\emph{(a)} $\Lambda$ is the cone with apex $q$ over a disk of radius $a\tan\alpha\le\frac35\bar R$ in $\{y_1=0\}$, so its points have norm at most $\frac65\bar R$. For $y$ in the slab, $-\bar R/2\le y_1\le0$ and $\tan\alpha\le\frac12$ give
\begin{align*}
    \|y'\|
    &\ \le\ (a-y_1)\tan\alpha
    \ \le\ \tfrac12\bigl(\tfrac65+\tfrac12\bigr)\bar R
    \ =\ \tfrac{17}{20}\bar R,\\
    \|y\|^2
    &\ \le\ \bigl(\tfrac14+\tfrac{289}{400}\bigr)\bar R^2
    \ \le\ \bar R^2 .
\end{align*}

\emph{(b)} Since $\sin\alpha=\bar R/a$, $q+K$ is the union of the rays from $q$ through $\cB_{\bar R}$; it is convex and contains $\cB_{\bar R}$ and $-q$, hence $\Omega$. As $0\in\cB_{\bar R}$ is the midpoint of $\pm q$, every point of $\Omega$ lies in $\conv(\cB_{\bar R}\cup\{q\})$ or in $\conv(\cB_{\bar R}\cup\{-q\})$: in $y=\lambda_1z+\lambda_2q-\lambda_3q$ with $z\in\cB_{\bar R}$ and $\lambda_2\ge\lambda_3$, say, replace $\lambda_3(q-q)$ by $2\lambda_3\cdot0$. Let $y\in\conv(\cB_{\bar R}\cup\{q\})\setminus\cB_{\bar R}$. If $y_1<b$, then $y_1<\|y\|\sin\alpha$, as $\|y\|>\bar R$, and $e:=y/\|y\|$ satisfies
\begin{align*}
    \langle q,e\rangle\ =\ \frac{a\,y_1}{\|y\|}\ <\ a\sin\alpha\ =\ \bar R,
\end{align*}
so the half-space $\{z:\langle z,e\rangle\le\bar R\}$ contains $\cB_{\bar R}$ and $q$ but not $y$, which is impossible; hence $y\in\Lambda$. Conversely, $\Lambda$ is the cone with apex $q$ over a disk of radius $(a-b)\tan\alpha=\bar R\cos\alpha$ in $\{y_1=b\}$, which lies in $\cB_{\bar R}$, so $\Lambda\subseteq\Omega$; by symmetry, $\Omega=\cB_{\bar R}\cup\Lambda\cup(-\Lambda)$. This cone has height $a-b=\bar R\cos^2\alpha/\sin\alpha$, so
\begin{align*}
    |\Lambda|=\frac{\omega_{n-1}}{n}\,\bar R^n\,\frac{\cos^{n+1}\alpha}{\sin\alpha}\ \le\ \frac{8\,\omega_{n-1}}{5n}\,\bar R^n\ \le\ \frac45\,\omega_n\bar R^n,
\end{align*}
because, for $\tan\alpha>\frac12$ and $n\ge2$,
\begin{align*}
    \frac{\cos^{n+1}\alpha}{\sin\alpha}
    \ \le\ \frac{\cos^3\alpha}{\sin\alpha}
    \ <\ \Bigl(\frac{2}{\sqrt5}\Bigr)^3\sqrt5
    \ =\ \frac85
\end{align*}
($\cos^3\alpha/\sin\alpha$ decreases in $\alpha$, and $\tan\alpha=\frac12$ gives $\cos\alpha=2/\sqrt5$, $\sin\alpha=1/\sqrt5$), and $n\omega_n\ge2\omega_{n-1}$, as each half of the unit sphere projects onto the unit ball of $\mR^{n-1}$. Hence
\begin{align*}
    |\Omega|\ \le\ \omega_n\bar R^n+2|\Lambda|\ \le\ \tfrac{13}{5}\,\omega_n\bar R^n ,
\end{align*}
which proves the conclusion.
\end{proof}

\begin{lemma}[Two-tip instances]
\label{lem:two-tip}
Let $R>0$, $0<\varsigma\le R/8$ and $\alpha\in(0,\pi/4]$, and choose $\bar R$ with $|\Omega|=\omega_nR^n+\vartheta_n(\alpha)\,\omega_n\varsigma^n$. Then for $i=0,1$:
(i) $S^{(i)}$ is compact, $|\partial S^{(i)}|=0$ and $|S^{(i)}|=\omega_nR^n$;
(ii) $W_0\cap W_1=\varnothing$ and $|W_i|=\vartheta_n(\alpha)\,\omega_n\varsigma^n$;
(iii) $S^{(i)}$ satisfies $\UIC(\alpha,3R/32)$, hence it is $(\vartheta_n(\alpha),3R/32)$-standard;
(iv) $d_H(S^{(0)},S^{(1)})\ge\varsigma$.
\end{lemma}

\begin{proof}
As $\Omega$ scales with $\bar R$, the volume condition determines $\bar R$, and $\omega_nR^n\le|\Omega|$ together with Lemma~\ref{lem:tip-body} gives $\bar R\ge\frac56R$ for thin tips and $\bar R\ge(\frac5{13})^{1/n}R>0.62R$ for wide tips.

\emph{(ii)} By symmetry consider $q_0=q$. For thin tips, $\cB_\varsigma(q)\subseteq\{y_1\ge a-\varsigma\}$ with $a-\varsigma>\bar R$, since
\begin{align*}
    a-\bar R\ =\ \frac{\bar R}{5}\ \ge\ \frac R6\ >\ \frac R8\ \ge\ \varsigma;
\end{align*}
so it misses $\cB_{\bar R}$ and $-\Lambda$ and meets $\Omega$ in $(q+K)\cap\cB_\varsigma(q)$. For wide tips, $\cB_\varsigma(q)\subseteq\{y_1\ge b\}$ because
\begin{align*}
    \varsigma\ \le\ \frac R8\ <\ \frac{\bar R}{\sqrt2}\ \le\ \frac{\bar R\cos^2\alpha}{\sin\alpha}\ =\ a-b,
\end{align*}
and there $\Omega$ coincides with $q+K$, since $\Omega\subseteq q+K$ and $\Lambda\subseteq\Omega$. In both cases $W_0=q+(K\cap\cB_\varsigma)$, of volume $\vartheta_n(\alpha)\,\omega_n\varsigma^n$, and $W_0\cap W_1=\varnothing$ because $2a>2\varsigma$.

\emph{(i) and (iv).} $\Omega$ is a finite union of convex bodies, so $S^{(i)}=\Omega\setminus\mathring{\cB}_\varsigma(q_{1-i})$ is compact with $\partial S^{(i)}\subseteq\partial\Omega\cup\partial\cB_\varsigma(q_{1-i})$, and, by (ii), $|S^{(i)}|=|\Omega|-|W_{1-i}|=\omega_nR^n$. For (iv), $q_0\in S^{(0)}$ while $S^{(1)}$ misses $\mathring{\cB}_\varsigma(q_0)$.

\emph{(iii)} By symmetry let $i=1$, so that $S^{(1)}$ cuts off the tip at $q_0$, and let $h:=3R/32$. We find for every $x\in S^{(1)}$ a unit vector $u$ with $x+(K_\alpha(u)\cap\cB_h)\subseteq S^{(1)}$. If $x\in q_0+K$ and $x+(K\cap\cB_h)\subseteq\Omega$, then $u:=-e_1$ works, because $\|x-q_0\|\ge\varsigma$ and Lemma~\ref{lem:cut-tip} keeps $x+K$ outside $\mathring{\cB}_\varsigma(q_0)$.

\emph{Thin tips.} Here $\bar R\ge\frac56R$ and $\alpha\le\pi/4$, so $\bar R\cos\alpha\ge\frac{5R}{6\sqrt2}>h$ and $\bar R/2>h$. If $x\in\cB_{\bar R}$, Proposition~\ref{prop:reach-implies-uic} gives a cone of height $\bar R\cos\alpha\ge h$ in $\cB_{\bar R}$, which misses $\mathring{\cB}_\varsigma(q_0)$. If $x\in\Lambda$ and $w\in K\cap\cB_h$, then $x+w\in q_0+K$ and $(x+w)_1\ge-h\ge-\bar R/2$, so $x+w\in\Lambda$ or, by Lemma~\ref{lem:tip-body}(a), $x+w\in\cB_{\bar R}$; hence $u:=-e_1$ works. If $x\in-\Lambda$, the mirror argument gives a cone in $\cB_{\bar R}\cup(-\Lambda)\subseteq\{y_1\le\bar R\}$, which misses $\mathring{\cB}_\varsigma(q_0)$.

\emph{Wide tips.} If $x_1\ge b+h$, then $x+w\in q_0+K$ and $(x+w)_1\ge b$ for $w\in K\cap\cB_h$, so $x+w\in\Lambda$ and $u:=-e_1$ works. If $x_1\le-(b+h)$, the mirror argument gives a cone in $-\Lambda\subseteq\{y_1<0\}$. If $|x_1|<b+h$, then $\|x\|\le\bar R+h$: either $x\in\cB_{\bar R}$, or $x\in\pm\Lambda$ with $|x_1|=b+\tau$, $\tau\in[0,h)$, where the cross-section of $\Lambda$ has radius at most $\bar R\cos\alpha$, so, as $b^2+\bar R^2\cos^2\alpha=\bar R^2$ and $b\le\bar R$,
\begin{align*}
    \|x\|^2
    &\ \le\ (b+\tau)^2+\bar R^2\cos^2\alpha\\
    &\ =\ \bar R^2+2b\tau+\tau^2\ \le\ (\bar R+\tau)^2 .
\end{align*} Let $u:=-x/\|x\|$ (any unit vector if $x=0$) and let $v$ be a unit vector at angle $\phi\le\alpha$ from $u$. Since
\begin{align*}
    \|x\|\sin\phi\ \le\ \frac{\bar R+h}{\sqrt2}\ <\ \bar R,
\end{align*}
the ray $x+tv$, $t\ge0$, meets $\cB_{\bar R}$ and leaves it at
\begin{align*}
    t_{\rm out}=\|x\|\cos\phi+\bigl(\bar R^2-\|x\|^2\sin^2\phi\bigr)^{1/2}\ \ge\ \bar R\cos\alpha\ \ge\ h;
\end{align*}
the first term is at least $\bar R\cos\alpha$ if $\|x\|\ge\bar R$, the second if $\|x\|<\bar R$. By convexity, $x+tv\in\Omega$ for $t\in[0,t_{\rm out}]$, so $x+(K_\alpha(u)\cap\cB_h)\subseteq\Omega$, and its points have $|y_1|<b+2h\le a-\varsigma$, so they miss $\mathring{\cB}_\varsigma(q_0)$. Here $\bar R>0.62R$ gives
\begin{align*}
    h&\ <\ (\sqrt2-1)\,\bar R,\qquad h\ \le\ \bar R\cos\alpha,\\
    \varsigma+2h&\ \le\ \tfrac{5}{16}R\ <\ \bar R/\sqrt2\ \le\ a-b . \qedhere
\end{align*}
\end{proof}

\begin{proof}[Proof of Theorem~\ref{thm:lower-std}]
The function $\vartheta_n$ is continuous and strictly increasing with $\vartheta_n(0^+)=0$ (Lemma~\ref{lem:circular-solid-angle}) and $\vartheta_n(\pi/4)=\kappa_n$, so there is a unique $\alpha\in(0,\pi/4]$ with $\vartheta_n(\alpha)=\kappa$.

Fix $\varsigma\in(2re^{-\mu T},R/8]$, chosen below independently of the estimator, let $S^{(0)}$, $S^{(1)}$, $W_0$ and $W_1$ be \latestrev{the sets of Lemma~\ref{lem:two-tip} translated by $c$}, and let $C:=S^{(0)}\cap S^{(1)}$, so that, up to null sets, $S^{(i)}$ is the disjoint union of $C$ and $W_i$.

\latestrev{The body-radius estimates in Lemma~\ref{lem:two-tip} give $\bar R\le R(1+\kappa8^{-n})^{1/n}$ and, before translation, $\Omega\subseteq\cB_{\sqrt5\bar R}\subseteq\cB_{3R}$. Set $F(x):=\mu(x-c)$ on $\X$. Then $\varphi(t,x)=c+e^{\mu t}(x-c)$ and, for $i=0,1$ and $t\in[0,T]$, $\varphi(t,S^{(i)})\subseteq\cB_{3R\max\{1,e^{\mu T}\}}(c)\subseteq\X$. Thus both instances satisfy Assumption~\ref{ass:well-posed}.}

Let
\begin{align*}
    p&:=\rho\,\kappa\,(\varsigma/R)^n\ \le\ \tfrac12,\\
    P_0^{(i)}&:=(1-p)\,\Unif(C)+p\,\Unif(W_i).
\end{align*}

By Lemma~\ref{lem:two-tip}(i) and~(iii), $S^{(i)}$ is compact with $|S^{(i)}|=\omega_nR^n$ and null boundary, and it is $(\kappa,3R/32)$-standard; standardness persists when $\lambda$ decreases. The law $P_0^{(i)}$ has density $p/|W_i|=\rho/|S^{(i)}|$ on $W_i$ and, as $|C|=\omega_nR^n(1-\kappa(\varsigma/R)^n)$ and $\rho\le1$, density
\begin{align*}
    \frac{1-p}{|C|}=\frac{1-\rho\,\kappa\,(\varsigma/R)^n}{1-\kappa\,(\varsigma/R)^n}\cdot\frac{1}{\omega_nR^n}\ \ge\ \frac{1}{|S^{(i)}|}\ \ge\ \frac{\rho}{|S^{(i)}|}
\end{align*}
on $C$. Thus $P_0^{(i)}$ is absolutely continuous, and its density is positive on $S^{(i)}$. By standardness, every ball centered in $S^{(i)}$ meets $S^{(i)}$ in positive volume, so the support of $P_0^{(i)}$ is $S^{(i)}$. The law $P_0^{(i)}$ also satisfies the density bound of Definition~\ref{def:std-family}, and it is the uniform law if $\rho=1$. Hence each instance lies in $\Fam^{\rm std}_{R,\mu,\kappa,\lambda,\rho}$ whenever $\lambda\le3R/32$.

By Lemma~\ref{lem:two-tip} (iv),
\begin{align*}
    d_H\bigl(S^{(0)}_T,S^{(1)}_T\bigr)\ \ge\ e^{\mu T}\varsigma\ >\ 2r,
\end{align*}
so, by the triangle inequality for Hausdorff distance, no estimate is within $r$ of both targets.

Draw independent $B_j\sim\mathrm{Bernoulli}(p)$, $U_j\sim\Unif(C)$ and $V_j^{(i)}\sim\Unif(W_i)$, and set $X_j^{(i)}:=U_j$ if $B_j=0$ and $X_j^{(i)}:=V_j^{(i)}$ otherwise. Each sequence is i.i.d.\ with law $P_0^{(i)}$. Let $\mathbb P$ denote the joint coupling law, and $P^{(i)}$ the marginal law of the endpoint data in instance $i$. On $E:=\{B_1=\dots=B_N=0\}$ the endpoint samples coincide, and so do the estimates. Let $A_i$ be the event that the estimate computed from instance $i$ has Hausdorff error greater than $r$. The separation of the targets gives $E\subseteq A_0\cup A_1$. Since $\log(1-p)\ge-2p$ for $p\le\frac12$,
\begin{align*}
    \max_iP^{(i)}(A_i)\ \ge\ \tfrac12\,\mathbb P(E)=\tfrac12(1-p)^N\ \ge\ \tfrac12e^{-2Np}.
\end{align*}

At $\varsigma=2re^{-\mu T}$,
\begin{align*}
    \frac1{2p}\log\frac1{2\delta}
    &\ =\ \frac{1}{2\rho\kappa}\Bigl(\frac{e^{\mu T}R}{2r}\Bigr)^n\log\frac1{2\delta}\\
    &\ =\ N_{\mathrm{lb}}(r,\delta)\ >\ N .
\end{align*} By continuity some $\varsigma\in(2re^{-\mu T},R/8]$ still has $2Np<\log\frac1{2\delta}$, and then $\max_iP^{(i)}(A_i)>\delta$.
The same argument covers randomized estimators by giving both copies the same random seed, independent of the coupled data.
\end{proof}

\begin{remark}[Range of $\kappa$]
\label{rem:lower-kappa}
Lemma~\ref{lem:cut-tip} needs $\alpha\le\pi/4$, which is why Theorem~\ref{thm:lower-std} stops at $\kappa_n=\vartheta_n(\pi/4)$. The range contains $(0,2^{-n}]$: $\kappa_2=\frac14$, $\kappa_3\approx0.146$ and $\kappa_4\approx0.091$ against $2^{-n}=0.25$, $0.125$ and $0.0625$. For $n\ge4$, bounding $\sin^{n-2}\phi\ge\sin^{n-2}\phi\cos\phi$ in \eqref{eq:vartheta-formula} gives
\begin{align*}
    \kappa_n
    &\ \ge\ \frac{(n-1)\,\omega_{n-1}}{n\,\omega_n}\int_0^{\pi/4}\sin^{n-2}\phi\cos\phi\,d\phi\\
    &\ =\ \frac{\omega_{n-1}}{n\,\omega_n}\,2^{-(n-1)/2}.
\end{align*}
Let $W_m:=\int_0^{\pi/2}\sin^m\phi\,d\phi$; it decreases in $m$ and $W_{m-1}W_m=\frac{\pi}{2m}$, so $W_m\le(\pi/(2m))^{1/2}$. With $\vartheta_n(\pi/2)=\frac12$ in \eqref{eq:vartheta-formula},
\begin{align*}
    \frac{n\,\omega_n}{\omega_{n-1}}
    &\ =\ 2(n-1)\,W_{n-2}
    \ \le\ 2(n-1)\Bigl(\frac{\pi}{2(n-2)}\Bigr)^{1/2}\\
    &\ \le\ 2^{(n+1)/2},
\end{align*}
where the last inequality holds at $n=4$ and the ratio of consecutive left-hand sides is at most $\frac43<\sqrt2$. Hence $\kappa_n\ge2^{-(n-1)/2}\,2^{-(n+1)/2}=2^{-n}$.
\end{remark}

\subsection{Standardness of the Main Classes of Sets}
\label{app:standard-examples}

This subsection proves Proposition~\ref{prop:standard-examples}, which the main text and the experiments use; the lower-bound proof does not.

A convex polytope with nonempty interior is written irredundantly as
\begin{equation}
    P=\{z:\ D_j(z):=\langle\nu_j,z\rangle-b_j\ge0,\ j=1,\dots,m\},
    \label{eq:polytope}
\end{equation}
with unit normals $\nu_j$, facets $F_j:=P\cap\{D_j=0\}$, vertex set $V(P)$ and vertex cones $T_v:=\{w:\langle\nu_j,w\rangle\ge0\text{ whenever }D_j(v)=0\}$, $v\in V(P)$.

\begin{proposition}[Classes of standard sets]
\label{prop:standard-examples}
Let $S_0\subset\mR^n$, $n\ge2$, be compact.

(i) If $S_0$ is the closure of an open set $A$ with $\reach(A^c)\ge r_0>0$, then $S_0$ is $(2^{-n},r_0)$-standard.

(ii) If $S_0$ satisfies $\UIC(\alpha,h)$ (Definition~\ref{def:uic}), then $S_0$ is $(\vartheta_n(\alpha),h)$-standard, where
\[
    \vartheta_n(\alpha)\ \ge\ \Bigl(\frac{\sin\alpha}{1+\sin\alpha}\Bigr)^n,
    \qquad
    \vartheta_2(\alpha)=\frac\alpha\pi .
\]

(iii) If $S_0=P$ is a convex polytope as in \eqref{eq:polytope}, then $S_0$ is $(\kappa_P,h_P)$-standard with $\kappa_P:=\min_{v\in V(P)}\Theta(T_v)$ and, with $J$ ranging over the nonempty subsets of $\{1,\dots,m\}$,
\begin{equation}
    h_P:=\min\Bigl\{\min_{z\in P}\max_{j\in J}D_j(z):\ \textstyle\bigcap_{j\in J}F_j=\varnothing\Bigr\}>0 .
    \label{eq:hP}
\end{equation}
The constant is optimal: $|\cB_r(v_\ast)\cap P|=\kappa_P\omega_nr^n$ for a minimizing vertex $v_\ast$ and all small $r$. If $S_0=\bigcup_{\ell\le k}P_\ell$ for convex polytopes $P_\ell$ with nonempty interior, then $S_0$ is $(\min_\ell\kappa_{P_\ell},\min_\ell h_{P_\ell})$-standard and $|\partial S_0|=0$.
\end{proposition}

\paragraph{Positive reach} Item~(i) is the following lemma.

\begin{lemma}[Positive reach gives local volume]
\label{lem:reach-volume}
If $S_0$ is compact and is the closure of an open set $A$ with $\reach(A^c)\ge r_0>0$, then $|\cB_r(x)\cap S_0|\ge2^{-n}\omega_nr^n$ for all $x\in S_0$ and $r\in(0,r_0]$.
\end{lemma}

\begin{proof}
First, every $x\in S_0$ lies in a closed ball $\cB_{r_0}(c)\subseteq S_0$. To see this for $x\in A$, put $C=A^c$. If $d(x,C)\ge r_0$, take $c=x$. Otherwise, let $a$ be the unique nearest point of $C$ and $u=(x-a)/d(x,C)$. The normal-ray property of positive reach \cite[Thm.~4.8(6)]{federer1959curvature} gives $d(a+tu,C)=t$ for $0<t<r_0$. By continuity, $c=a+r_0u$ has $d(c,C)=r_0$, so $\cB_{r_0}(c)\subseteq\overline A$ and contains $x$. For $x\in\partial A$, approximate $x$ by points of $A$ and take a convergent subsequence of the corresponding centers, which lie in the compact set $S_0$. Let $d:=\|x-c\|$ and $r\le r_0$. It suffices to find a ball of radius $r/2$ inside $\cB_r(x)\cap\cB_{r_0}(c)$. If $d<r/2$, the ball $\cB_{r/2}(c)$ works. If $d\ge r/2$, let $q:=x+\frac r2\frac{c-x}{d}$; for $z\in\cB_{r/2}(q)$,
\begin{align*}
    \|z-x\|&\le\|z-q\|+\|q-x\|\le r,\\
    \|z-c\|&\le\|z-q\|+\|q-c\|\le\tfrac r2+\bigl(d-\tfrac r2\bigr)=d\le r_0 . \qedhere
\end{align*}
\end{proof}

\paragraph{Polytopes} For a polytope $P$ as in \eqref{eq:polytope} and $x\in P$, let $J(x):=\{j:D_j(x)=0\}$ be the active set, $T_x:=\{w:\langle\nu_j,w\rangle\ge0\ \forall j\in J(x)\}$ the tangent cone, which is the vertex cone $T_v$ when $x=v\in V(P)$, and $s_x:=\min_{j\notin J(x)}D_j(x)>0$. For $s\le s_x$, $P\cap\cB_s(x)=(x+T_x)\cap\cB_s(x)$: each $D_j$ is $1$-Lipschitz, so $D_j(z)\ge s_x-s\ge0$ for $j\notin J(x)$, while $D_j(z)=\langle\nu_j,z-x\rangle$ for $j\in J(x)$.

\begin{theorem}[Convex polytopes satisfy a uniform vertex-cone condition]
\label{thm:polytope-uic}
Let $P$ be as in \eqref{eq:polytope}, $S_0:=P$, $\cK_P:=\{T_v:v\in V(P)\}$ and $h_P$ as in \eqref{eq:hP}. Then $h_P>0$, and every $x\in P$ has a vertex $v$ with $x+(T_v\cap\cB_{h_P})\subseteq P$. Hence $S_0$ satisfies $\UIC(\cK_P,h_P)$ and is $(\kappa_P,h_P)$-standard with $\kappa_P=\min_{v\in V(P)}\Theta(T_v)$, and $|\cB_r(v_\ast)\cap P|=\kappa_P\omega_nr^n$ for a minimizing vertex $v_\ast$ and $r\le s_{v_\ast}$.
\end{theorem}

\begin{proof}
\emph{$h_P>0$.} The family of index sets in \eqref{eq:hP} is finite and nonempty: if $D_j(z)=0$ for all $j$ and $p\in\inte P$, then $D_j(z+t(p-z))=tD_j(p)\ge0$ for all $t\ge0$, and $P$ would contain a ray. For each admissible $J$, $\max_{j\in J}D_j$ attains its minimum on $P$, and the minimum is positive, since a zero would lie in $\bigcap_{j\in J}F_j=\varnothing$.

\emph{Vertex cones.} Fix $x\in P$ and $J:=\{j:D_j(x)<h_P\}$. If $J=\varnothing$, choose any vertex $v$. Otherwise $\bigcap_{j\in J}F_j\ne\varnothing$, since otherwise $h_P\le\max_{j\in J}D_j(x)<h_P$. In the latter case, this face contains a vertex $v$ with $J\subseteq J(v)$. For $w\in T_v\cap\cB_{h_P}$, $D_j(x+w)\ge D_j(x)\ge0$ for $j\in J$ and $D_j(x+w)\ge D_j(x)-\|w\|\ge0$ for $j\notin J$; hence $x+w\in P$. For $r\le s_{v_\ast}$, the local conical identity gives $|\cB_r(v_\ast)\cap P|=|T_{v_\ast}\cap\cB_r|=\kappa_P\omega_nr^n$.
\end{proof}

\begin{corollary}[Simplices and finite unions]
\label{cor:simplex}
(i) For a nondegenerate simplex $\Delta$ with heights $\eta_i$, $h_\Delta$ equals the inradius $\varrho_\Delta=(\sum_i\eta_i^{-1})^{-1}$, so $\Delta$ is $(\min_i\Theta(T_{v_i}),\varrho_\Delta)$-standard.
(ii) If $S_0=\bigcup_{\ell\le k}P_\ell$ for convex polytopes $P_\ell$ with nonempty interior, then $S_0$ is $(\min_\ell\kappa_{P_\ell},\min_\ell h_{P_\ell})$-standard and $|\partial S_0|=0$.
\end{corollary}

\begin{proof}
\emph{(i)} With barycentric coordinates $\lambda_i$, $D_i=\eta_i\lambda_i$ and $\sum_iD_i/\eta_i=1$, so $\max_iD_i\ge\varrho_\Delta$, with equality at the incenter. For a simplex, $\bigcap_{j\in J}F_j=\conv\{v_k:k\notin J\}$ is empty only for $J=\{0,\dots,n\}$, so \eqref{eq:hP} reduces to $h_\Delta=\min_z\max_iD_i(z)=\varrho_\Delta$; apply Theorem~\ref{thm:polytope-uic}.
\emph{(ii)} Every $x\in S_0$ lies in some $P_\ell$; Theorem~\ref{thm:polytope-uic} gives a cone $T$ with $\Theta(T)\ge\kappa_{P_\ell}$ and $x+(T\cap\cB_{h_{P_\ell}})\subseteq P_\ell\subseteq S_0$. Apply Lemma~\ref{lem:uic-standard}; $\partial S_0\subseteq\bigcup_\ell\partial P_\ell$ is null. Reflex vertices are therefore harmless: triangulate and apply (ii).
\end{proof}

\begin{proof}[Proof of Proposition~\ref{prop:standard-examples}]
For (i), apply Lemma~\ref{lem:reach-volume} with $\kappa=2^{-n}$ and $\lambda=r_0$. For (ii), Lemma~\ref{lem:circular-solid-angle} bounds $\vartheta_n(\alpha)$ below, and Lemma~\ref{lem:uic-standard} converts $\UIC(\alpha,h)$ into $(\vartheta_n(\alpha),h)$-standardness. For (iii), Theorem~\ref{thm:polytope-uic} supplies $(\kappa_P,h_P)$, and Corollary~\ref{cor:simplex}(ii) extends the constants to finite unions with $|\partial S_0|=0$.
\end{proof}

\begin{IEEEbiography}[{\includegraphics[width=1in,height=1.25in,clip,keepaspectratio]{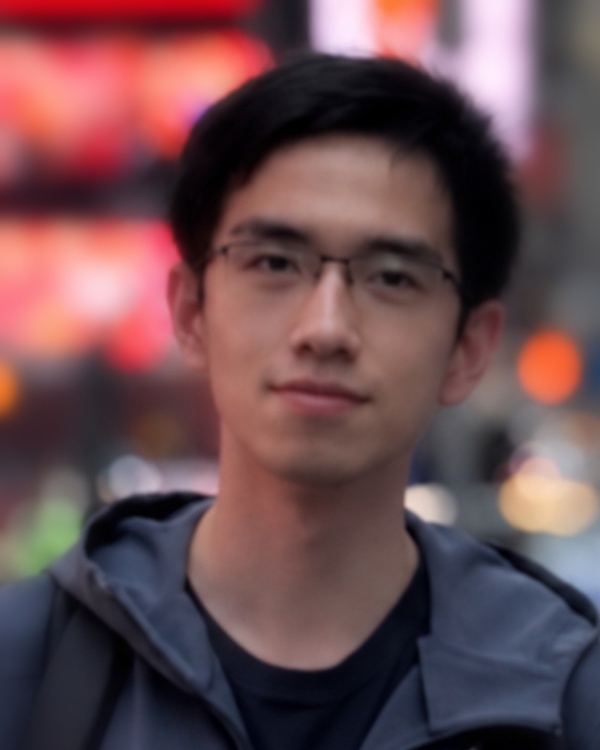}}]{Jixian Liu}
is currently pursuing the Ph.D. degree with the Data Science and AI Institute, Johns Hopkins University, Baltimore, MD, USA. He received the B.Sc. degree from the School of Physics, Xi'an Jiaotong University, Xi'an, Shaanxi, China, in 2024. His research interests include robotics, control, and verifiable autonomous systems, with a focus on data-driven, learning-based and generative methods for safety verification.
\end{IEEEbiography}

\begin{IEEEbiography}[{\includegraphics[width=1in,height=1.25in,clip,keepaspectratio]{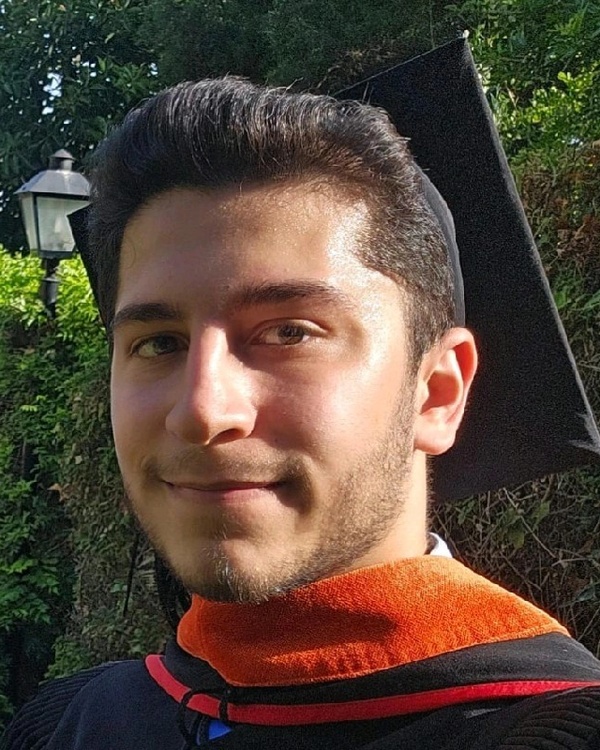}}]{Ihab Tabbara}
is currently pursuing the Ph.D. degree in Computer Science and Engineering at Washington University in St. Louis, St. Louis, MO, USA. He received the B.E. degree in Computer and Communication Engineering from the American University of Beirut, Beirut, Lebanon, in 2024. His research interests include robotics, machine learning, safe control, and formal methods.
\end{IEEEbiography}

\begin{IEEEbiography}[{\includegraphics[width=1in,height=1.25in,clip,keepaspectratio]{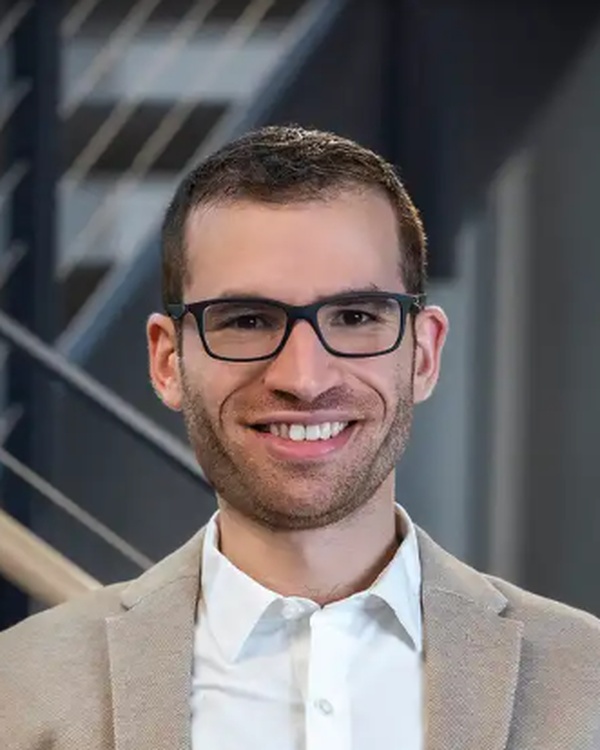}}]{Hussein Sibai}
is an Assistant Professor in the Department of Computer Science and Engineering at Washington University in St.\ Louis, St.\ Louis, MO, USA. He was previously a postdoctoral scholar at the University of California, Berkeley. He received the bachelor's degree in computer and communication engineering from the American University of Beirut, Beirut, Lebanon, and the master's and Ph.D. degrees in electrical and computer engineering from the University of Illinois Urbana-Champaign, Urbana, IL, USA, completing his Ph.D. in December 2021. His research interests include robotics, formal methods, control theory, and reinforcement learning. He received the Best Poster Award at HSCC 2018 and best paper nominations at HSCC 2017 and ATVA 2019. His honors include the Rambus Fellowship, the Ernest A.\ Reid Fellowship, the MAVIS Future Faculty Fellowship, and the ACM SIGBED gold medal in the graduate Student Research Competition at CPS Week 2021.
\end{IEEEbiography}

\begin{IEEEbiography}[{\includegraphics[width=1in,height=1.25in,clip,keepaspectratio]{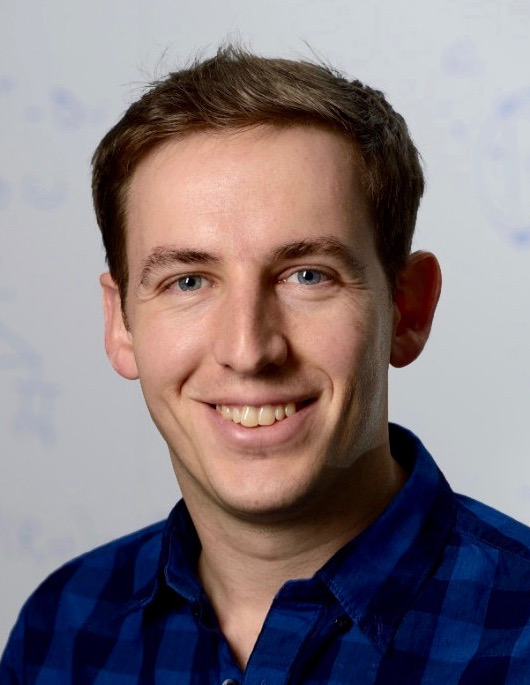}}]{Enrique Mallada}
(S'09-M'13-SM'19) is an Associate Professor of Electrical and Computer Engineering at Johns Hopkins University. Prior to joining Hopkins in 2016, he was a Postdoctoral Fellow in the Center for the Mathematics of Information at Caltech from 2014 to 2016. He received his Ingeniero en Telecomunicaciones degree from Universidad ORT, Uruguay, in 2005 and his Ph.D. degree in Electrical and Computer Engineering with a minor in Applied Mathematics from Cornell University in 2014.
Dr. Mallada was awarded
the NSF CAREER award in 2018,
the ECE Director's PhD Thesis Research Award for his dissertation in 2014,
the Center for the Mathematics of Information (CMI) Fellowship from Caltech in 2014,
and the Cornell University Jacobs Fellowship in 2011.
His research interests lie in the areas of control, dynamical systems and optimization, with applications to engineering networks such as power systems and the Internet.
\end{IEEEbiography}

\end{document}